\documentclass[12pt, a4paper] {article}

\usepackage{amsfonts}
\usepackage{color}
\usepackage{amsmath}
\usepackage{graphicx}
\usepackage{bm}
\usepackage{amssymb}
\usepackage{latexsym}
\usepackage{mathrsfs}
\usepackage{textcomp}
\begin{document}

\begin{center}

{\sf {\bf \large  Nilpotent Symmetries in Jackiw$-$Pi Model: Augmented Superfield Approach}}

\vskip 1.0 cm

{\sf{ \bf Saurabh Gupta$^{(a)}$ and R. Kumar$^{(b)}$}}\\
\vskip .1cm
{\it $^{(a)}$Instituto de F\'\i sica, Universidade de S\~ao Paulo,\\
C. Postal 66318,  05314-970 S\~ao Paulo, SP, Brazil}\\
\vskip .3cm
{\it $^{(b)}$S. N. Bose National Centre for Basic Sciences,\\
Block JD, Sector III, Salt Lake, Kolkata$-$700098, India}\\
\vskip .1cm
{\sf E-mails: saurabh@if.usp.br; rohit.kumar@bose.res.in}

\end{center}

\vskip 1.0 cm

\noindent

\noindent
{\bf Abstract:} We derive the complete set of off-shell nilpotent ($s^2_{(a)b} = 0$) and absolutely anticommuting 
($s_b s_{ab} + s_{ab} s_b = 0$) Becchi-Rouet-Stora-Tyutin (BRST) ($s_b$) as well as anti-BRST symmetry 
transformations ($s_{ab}$) corresponding to the combined Yang-Mills and non-Yang-Mills symmetries 
of the $(2 + 1)$-dimensional Jackiw-Pi model within the framework of augmented superfield formalism. 
The absolute anticommutativity of the (anti-)BRST symmetries is ensured by the existence of {\it two} sets of Curci-Ferrari (CF) type of 
conditions which emerge naturally in this formalism. The presence of CF conditions enables us to derive the coupled but 
equivalent Lagrangian densities. We also capture the (anti-)BRST invariance of the coupled Lagrangian densities 
in the superfield formalism. The derivation of the (anti-)BRST transformations of the auxiliary field $\rho$ is one of the key findings which can 
neither be generated by the nilpotent (anti-)BRST charges nor by the requirements of the nilpotency and/or absolute 
anticommutativity of the (anti-)BRST transformations. Finally, we provide a bird's-eye view on the role of 
auxiliary field for various massive models and point out few striking similarities and some glaring differences among them.\\ 



\newpage

\section{Introduction}
The co-existence of mass and gauge invariance {\it together} is still one of the main issues connected with the gauge theories, 
in spite of the astonishing success of the standard model of particle physics which is based on (non-)Abelian 1-form gauge theories.
However, it is worthwhile to mention that, in the case of sufficiently strong vector couplings, the gauge invariance does not entail 
the masslessness of gauge particles \cite{Schwinger:1962tn,Schwinger:1962tp}. Thus, it is needless to say that the mass generation 
in gauge theories is a crucial issue which has attracted a great deal of interest \cite{Deser:1981wh,Deser:1982vy}.

In the recent past, many models for the mass generation have been studied in the diverse dimensions of spacetime. In this context,
mention can be made of about 4D topologically massive (non-)Abelian gauge theories, with $(B \wedge F)$ term, where 1-form gauge 
field acquires a mass in a natural fashion \cite{Freedman:1980us,Allen:1990gb,Harikumar:2001eb}. One of the key features associated 
with such models is that the 1-form gauge field gets a mass without taking any recourse to the Higgs mechanism. We have 
thoroughly investigated these models within the framework of Becchi-Rouet-Stora-Tyutin (BRST) as well as superfield formalism 
\cite{Gupta:2008he,Gupta:2010xh,Gupta:2009up,Kumar:2011zi,Krishna:2010dc,Malik:2011pm}. It is interesting to point out that the main issues 
connected with the 4D Abelian topologically massive models are that they suffer from the problems connected with renormalizability when
 straightforwardly generalized to the non-Abelian case \cite{Henneaux:1997mf}. However, this issue can be circumvented by the 
introduction of extra field (see, e.g. \cite{Lahiri:1996dm,Lahiri:1999uc}).

At this juncture, it is worth mentioning about the lower dimensional non-Abelian massive models, such as $(2 + 1)$-dimensional 
Jackiw-Pi (JP) model \cite{Jackiw:1997jga}, which are free from the above mentioned issues. The silent features of JP model 
are as follows. First, it is parity conserving model due to the introduction of a 1-form vector field having odd parity. 
Second, mass and gauge invariance are respected together. Third, it is endowed with the two independent sets of local continuous 
symmetries, namely; the usual Yang-Mills (YM) symmetries and non-Yang-Mills (NYM) symmetries. Finally, it is free from the problems 
connected with the 4D topologically massive models. These features make JP model attractive and worth studying in detail.

The JP model has been explored in many different prospects such as constraint analysis and Hamiltonian formalism \cite{Dayi:1997in}, 
establishment of Slavnov-Taylor identities and BRST symmetries \cite{DelCima:2011bx}. Furthermore, this model is also shown to be 
ultraviolet finite and renormalizable \cite{DelCima:2012bm}. We have applied superfield formalism and derived the full set of 
off-shell nilpotent and absolutely anticommuting BRST as well as anti-BRST symmetry transformations corresponding to the both YM 
and NYM symmetries of JP model \cite{Gupta:2011cta,Gupta:2012ur}. Within the superfield formalism, we have been able to derive the 
{\it proper} (anti-)BRST transformations for the auxiliary field $\rho$ which can neither be deduced by the conventional means of nilpotency 
and/or absolute anticommutativity of (anti-)BRST symmetries nor generated by the conserved (anti-)BRST charges. At this stage, we would like to 
point out that the derivation of the proper anti-BRST symmetries have utmost importance because they play a fundamental role in the BRST 
formalism (see, e.g. \cite{Curci:1976ar,Ojima:1980da,Hwang:1989mn} for details). In fact, both the symmetries (i.e. BRST and anti-BRST) 
have been formulated in an independent way \cite{Hwang:1983sm}.

Recently, the (anti-)BRST symmetries for perturbative quantum gravity in curved as well as complex spacetime, in linear 
as well as in non-linear gauges have been found \cite{mir1,mir2} and a superspace 
formulation of higher derivative theories \cite{mir3}, Chern-Simons and Yang-Mills theories on deformed 
superspace \cite{mir4,mir5}  within BV formalism have also been 
established. Moreover, the study of massless and massive fields with totally symmetric arbitrary spin in AdS space has 
been carried  out in the framework of BRST formalism \cite{mets}.

The main motivations behind our present investigation are as follows. First, the derivation of off-shell nilpotent and absolutely 
anticommuting (anti-)BRST symmetry transformations corresponding to the combined YM and NYM symmetries of JP model. As, in 
our recent works (cf. \cite{Gupta:2011cta,Gupta:2012ur}), we have already established the corresponding proper (anti-) BRST 
symmetry transformations, individually, for both the YM and NYM cases, within the framework of superfield formalism. Second, 
to establish the Curci-Ferrari (CF) conditions in the case of combined symmetries. These CF conditions are hallmark of 
any non-Abelian 1-form gauge theories \cite{Curci:1976ar} and have a close connection with gerbes \cite{Bonora:2007hw}, 
within the framework of BRST formalism. Third, to procure appropriate coupled Lagrangian densities which respect the 
(anti-)BRST symmetries derived from augmented superfield approach. Finally, to point out the role of auxiliary field 
$\rho$, which is very special to this model (cf. \cite{Dayi:1997in,Gupta:2011cta} for details).

This paper is organized in the following manner. In Section 2, we recapitulate the underlying symmetries of 3D JP model. 
We derive the off-shell nilpotent and absolutely anticommuting (anti-)BRST symmetries corresponding to the combined YM and 
NYM symmetries of JP model, within the framework of superfield formalism, in Section 3. Section 4 contains the derivation 
of coupled Lagrangian densities that respect the preceding (anti-) BRST symmetries. The conservation of (anti-)BRST charges is 
shown in Section 5. We also discuss about the novel observations of our present study in this section. Section 6 is 
devoted for the discussions of ghost symmetries and BRST algebra. In Section 7, we provide a bird's-eye view on the role 
of auxiliary field in the context of various massive models. Finally, in Section 8, we make some concluding remarks.

In Appendix A, we show the nilpotency and absolute anticommutativity of the (anti-) BRST charges within the framework
of augmented superfield formalism. We also capture (anti-)BRST invariance of coupled Lagrangian densities in 
the superfield framework.

{\sl Conventions and notation:} We adopt here the conventions and notation such that the 3D 
flat Minkowski metric $\eta_{\mu\nu} =$ diag $(+ 1, - 1, -1)$ and the 3D totally antisymmetric 
Levi-Civita tensor $\varepsilon_{\mu\nu\eta}$ satisfies $\varepsilon_{\mu\nu\eta}\,\varepsilon^{\mu\nu\eta}= - 3!,\; 
\varepsilon_{\mu\nu\eta}\,\varepsilon^{\mu\nu\kappa}$ $= - 2! \delta^\kappa_\eta$, etc. with $\varepsilon_{012} = - \varepsilon^{012} = +1$. 
The Greek indices $\mu, \nu, \eta, ... = 0, 1, 2$ correspond to the 3D spacetime directions and Latin indices $i, j, k,... = 1,2$ 
correspond to the space directions only. The dot and cross product between two non-null vectors $P$ and $Q$ 
in the $SU(N)$ Lie algebraic space are defined as $P \cdot Q = P^a Q^a,\; P \times Q = f^{abc}\, P^a Q^b T^c$. The $SU(N)$ generators
$T^a$ (with $a, b, c,... = N^2 - 1$) follow the commutation relation $[T^a,\, T^b] = i f^{abc}\, T^c$ where the structure constants 
$f^{abc}$ are chosen to be totally antisymmetric in $a,b,c$ for the semi-simple $SU(N)$ Lie algebra \cite{Wein}.

\section{Preliminaries:  Jackiw-Pi model}
We start off with the massive, non-Abelian, gauge invariant Jackiw-Pi model in $(2+ 1)$-dimensions 
of spacetime. The Lagrangian density of this model is given by \cite{Jackiw:1997jga,Gupta:2011cta}
\begin{eqnarray}
{\cal L}_0 &=& - \frac{1}{4}\, F_{\mu\nu}\cdot F^{\mu\nu} 
- \frac{1}{4}\, \big(G_{\mu\nu} + g F_{\mu\nu} \times \rho\big) \cdot \big(G^{\mu\nu} + g F^{\mu\nu} \times \rho\big) \nonumber\\
&+& \frac{m}{2}\, \varepsilon^{\mu\nu\eta}\,F_{\mu\nu}\cdot \phi_\eta, \label{2.1}
\end{eqnarray} 
where the 2-form $F^{(2)} = d A^{(1)} + i g \big(A^{(1)} \wedge A^{(1)} \big) 
= \frac{1}{2!}\,\big(dx^\mu \wedge dx^\nu \big)F_{\mu\nu}\cdot T$
defines the curvature tensor $F_{\mu\nu} = \partial_\mu A_\nu - \partial_\nu A_\mu - g(A_\mu \times A_\nu)$ 
for the non-Abelian 1-form [$A^{(1)} = dx_\mu A^\mu \cdot T$] gauge field $A_\mu = A_\mu \cdot T$
where $d = dx^\mu \partial_\mu$ is the exterior derivative (with $d^2 = 0$). Similarly, 
another 2-form $G^{(2)} = d\phi^{(1)} + i g \big(A^{(1)} \wedge \phi^{(1)}\big) + ig \big(\phi^{(1)} \wedge A^{(1)}\big) 
= \frac{1}{2!}\,\big(dx^\mu \wedge dx^\nu\big)\,G_{\mu\nu} \cdot T$ defines the curvature tensor 
$G_{\mu\nu} = D_\mu \phi_\nu - D_\nu \phi_\mu$ corresponding\footnote{The covariant derivative is defined 
as $D_\mu *  = \partial_\mu * - g (A_\mu \times * )$. } to 1-form $[\phi^{(1)} = dx^\mu\phi_\mu \cdot T]$ 
vector field $\phi_\mu = \phi_\mu \cdot T$. In the above, the vector fields $A_\mu$ and $\phi_\mu$ have opposite parity 
thus the JP model becomes parity invariant, $\rho$ is an auxiliary field, $g$ is the coupling constant and $m$ 
defines the mass parameter.

\subsection{Local gauge symmetries: YM and NYM}
The above Lagrangian density respects two sets of local and continuous gauge symmetry transformations, namely;  
YM gauge transformations $(\delta_1)$ and NYM gauge transformations $(\delta_2)$. These symmetry transformations 
are \cite{Gupta:2011cta,Gupta:2012ur}
\begin{eqnarray}
&&\delta_1 A_\mu = D_\mu \Lambda, \quad \delta_1 \phi_\mu = - g(\phi_\mu \times \Lambda), 
\quad \delta_1 \rho = - g(\rho \times \Lambda),\nonumber\\
&& \delta_1 F_{\mu\nu} = - g(F_{\mu\nu} \times \Lambda), 
\quad \delta_1 G_{\mu\nu} = - g(G_{\mu\nu} \times \Lambda), \label{2.2}
\end{eqnarray}
\begin{eqnarray}
\delta_2 A_\mu = 0, \quad \delta_2 \phi_\mu = D_\mu \Sigma, \quad \delta_2 \rho = + \Sigma, \quad
\delta_2 F_{\mu\nu} = 0, \quad \delta_2 G_{\mu\nu} = - g(F_{\mu\nu} \times \Sigma), \label{2.3}
\end{eqnarray}
where $\Lambda \equiv \Lambda \cdot T $ and $\Sigma \equiv \Sigma \cdot T$ are the $SU(N)$ valued local gauge parameters 
corresponding to the YM and NYM gauge transformations, respectively. Under the above  local and infinitesimal gauge 
transformations the Lagrangian density (\ref{2.1}) transforms as 
\begin{eqnarray}
&&\delta_1 {\cal L}_0 = 0, \qquad \delta_2 {\cal L}_0 = \partial_\mu \Big[\frac{m}{2}\, \varepsilon^{\mu\nu\eta}\, F_{\nu\eta}\cdot \Sigma\Big].  
\end{eqnarray}
As a consequence, the action integral $S = \int d^3x {\cal L}_0$ remains invariant under both the gauge 
transformations ($\delta_1$ and $\delta_2$) for the physically well-defined fields which vanish off rapidly 
at infinity. We would like to point out that in order to maintain the NYM symmetry, we have to have the 
auxiliary field $\rho$ in the theory (cf. Section 7 for details).

\subsection{Combined gauge symmetry}
In the above, we have seen that both the YM and NYM transformations are the symmetries of the theory.  
Thus, the combination of the above symmetries [i.e. $(\delta = \delta_1 + \delta_2$)] would also be the 
symmetry of theory. Under the combined gauge transformation $\delta$, namely;
\begin{eqnarray}
&&\delta A_\mu = D_\mu \Lambda, \qquad \delta \phi_\mu = D_\mu \Sigma - g(\phi_\mu \times \Lambda), \qquad 
\delta \rho = \Sigma - g(\rho \times \Lambda),\nonumber\\
&& \delta F_{\mu\nu} = - g(F_{\mu\nu} \times \Lambda), \qquad \delta G_{\mu\nu} = - g(G_{\mu\nu} \times \Lambda) 
-g (F_{\mu\nu} \times \Sigma), \label{2.5}
\end{eqnarray}
the Lagrangian density (\ref{2.1}) remains quasi-invariant. To be more specific, the Lagrangian density transforms 
to a total spacetime derivative 
\begin{eqnarray}
\delta {\cal L}_0 = \partial_\mu \Big[\frac{m}{2}\, \varepsilon^{\mu\nu\eta}\, F_{\nu\eta}\cdot \Sigma\Big]. 
\end{eqnarray}
Thus, the action integral remains invariant (i.e. $\delta S = \delta \int d^3x {\cal L}_0 = 0$) under the 
combined symmetry ($\delta$), too.

\section{(Augmented) superfield approach}
We apply Bonora-Tonin's (BT) superfield approach to the BRST formalism \cite{Bonora:1980pt,Bonora:1980ar}, 
to derive the off-shell nilpotent and absolutely anticommuting (anti-) BRST symmetry transformations for the 
1-form gauge field $A_\mu$ and corresponding (anti-)ghost fields $(\bar C)C$.

\subsection{(Anti-)BRST symmetries: Gauge and (anti-)ghost fields}
For this purpose, we generalize 1-form connection $A^{(1)}$ (and corresponding 2-form curvature $F^{(2)}$) and exterior 
derivative $d$ onto  the $(3,2)$-dimensional supermanifold, as
\begin{eqnarray}
d \to \tilde d &=& dZ^M\partial_M = dx^\mu\,\partial_\mu + d\theta\, \partial_\theta + d\bar\theta \,\partial_{\bar\theta}, 
\qquad \tilde d^2 = 0,\nonumber\\
A^{(1)} \to \tilde{\cal A}^{(1)} &=& dZ^M \tilde {\cal A}_M = dx^\mu\,\tilde {\cal A}_\mu(x, \theta, \bar\theta) 
+ d\theta\, {\tilde {\bar{ \cal F}}} (x, \theta, \bar\theta) 
+ d \bar\theta\, {\tilde{{\cal F}}} (x, \theta, \bar\theta), \nonumber\\
F^{(2)} \to \tilde {\cal F}^{(2)} &=& \frac{1}{2!}\,(dx^M \wedge dx^N)\,\tilde {\cal F}_{MN} = \tilde d \tilde {\cal A}^{(1)} 
+ i g \big(\tilde {\cal A}^{(1)} \wedge \tilde {\cal A}^{(1)} \big)  ,
\end{eqnarray}
where $Z^M = (x^\mu, \theta, \bar \theta)$ are superspace coordinates characterizing the $(3, 2)$-dimensional supermanifold. 
In the above expression, $\theta$ and $\bar \theta$ are the Grassmannian variables (with $\theta^2 = \bar \theta^2 = 0, 
\theta \bar \theta + \bar \theta \theta = 0$) and $\partial_\theta, \partial_{\bar\theta}$ are corresponding Grassmannian 
derivatives. We also generalize 3D gauge field [$A_\mu (x)$] and (anti-)ghost fields $[(\bar C)C(x)]$ of the theory to 
their corresponding superfields onto the $(3,2)$-dimensional supermanifold.

Now, these superfields can be expanded along 
the Grassmannian directions, in terms of the basic fields ($A_\mu, C, \bar C$) and secondary fields 
($R_\mu$, $\bar R_\mu$, $S_\mu$, $B_1$, $B_2$, $\bar B_1$, $\bar B_2$, $s$, $\bar s$), in the following manner, 
\begin{eqnarray}
\tilde {\cal A}_{\mu} (x, \theta, \bar\theta) &=& A_\mu (x) + \theta\, \bar R_\mu (x) + \bar \theta\, R_\mu (x)
+ i \,\theta \,\bar\theta \, S_\mu (x), \nonumber\\
\tilde {\cal F} (x, \theta, \bar\theta) &=& C (x) + i\,\theta\, \bar B_1 (x) + i\,\bar \theta\, B_1 (x)
+ i \,\theta\, \bar\theta \, s (x), \nonumber\\
{\tilde {\bar {\cal F}}} (x, \theta, \bar\theta) &=& \bar C (x) +  i\,\theta\, \bar B_2 (x) + i\,\bar \theta\; 
B_2 (x) + i \,\theta \,\bar\theta \; \bar s (x). \label{3.2}
\end{eqnarray}
Here $ \tilde {\cal A}_{\mu} (x, \theta, \bar\theta), \tilde {\cal F} (x, \theta, \bar\theta), {\tilde{\bar {\cal F}}} (x, \theta, \bar\theta) $ 
are superfields corresponding to the basic fields $A_\mu (x),$  $C (x)$ and $\bar C (x)$, respectively. 
Now these secondary fields, in the above expression, can be determined in terms of the basic and auxiliary fields of the underlying theory through 
the application of horizontality condition (HC) (cf. \cite{Bonora:1980pt,Bonora:1980ar} for details). This HC can be 
mathematically expressed in the following fashion
\begin{eqnarray}
d\,A^{(1)} + i \,g \big(A^{(1)} \wedge A^{(1)}\big)= \tilde d \,\tilde{\cal A}^{(1)} 
+ i \,g\big(\tilde{\cal A}^{(1)} \wedge \tilde{\cal A}^{(1)}\big) \Longleftrightarrow  F^{(2)} = \tilde {\cal F}^{(2)}. \label{3.3}
\end{eqnarray}
Exploiting the above HC, we obtain the following relationships among the basic, auxiliary and secondary fields of the theory
\begin{eqnarray}
&& R_\mu = D_\mu C, \quad \bar R_\mu = D_\mu \bar C, \quad B_1 = -  \frac{i}{2}\,g\,(C \times C),\quad
 \bar B_2 = - \frac{i}{2}\, g\,(\bar C \times \bar C), \nonumber\\
&& B + \bar B = -\,i\, g\,(C \times \bar C),\quad
s = -\,g\,(\bar B \times C), \quad \bar s = +\, g\,(B \times \bar C), \nonumber\\
&& S_\mu = D_\mu B \,+ \,i\, g \,(D_\mu C \times \bar C) \equiv  - \,D_\mu \bar B - \,i\, g\,(D_\mu \bar C \times C), \label{3.4}
\end{eqnarray}
where we have chosen $\bar B_1 = \bar B$ and $B_2 = B$.

Substituting the relationships (\ref{3.4}) into the super-expansion of superfields in (\ref{3.2}), we procure following explicit 
expansions
\begin{eqnarray}
\tilde {\cal A}^{(h)}_{\mu} (x, \theta, \bar\theta) &=& A_\mu (x) + \theta D_\mu \bar C (x) 
+ \bar \theta D_\mu C (x)+  \theta \,\bar\theta \, \big[i D_\mu  B - g(D_\mu C \times \bar C)\big](x) \nonumber\\
&\equiv& A_\mu (x) + \theta \big(s_{ab}\, A_\mu (x)\big) + \bar \theta \big(s_b \,A_\mu (x)\big) 
+ \theta \bar\theta \big(s_b\, s_{ab} \,A_\mu (x)\big), \nonumber\\
\tilde {\cal F}^{(h)} (x, \theta, \bar\theta) &=& C (x) + \theta \big(i \bar B (x)\big) + \bar \theta \Bigl 
[\frac{g}{2}\, (C \times C) \Bigr](x) + \theta \bar\theta  \big[-i g\big(\bar B \times C)\big](x) \nonumber\\
&\equiv& C (x) + \theta \big(s_{ab} C (x)\big) + \bar \theta \big(s_b C (x)\big) 
+ \theta \bar\theta \big(s_b s_{ab} \,C (x)\big),\nonumber\\
 {\tilde {\bar {\cal F}}}^{(h)} (x, \theta, \bar\theta) &=& \bar C (x) +
\theta \Bigl [\frac{g}{2} \,(\bar C \times \bar C)\Bigr](x) + \bar \theta \big(i B (x)\big)
+ \theta \bar\theta  \big[i g(B \times \bar C)\big](x) \nonumber\\
&\equiv& \bar C (x) + \theta \big(s_{ab} \bar C (x)\big) + \bar \theta\, \big(s_b \bar C (x)\big)
+ \theta \bar\theta \big(s_b \,s_{ab}\, \bar C (x)\big). \label{3.5}
\end{eqnarray}
In the above, the superscript $(h)$ on the superfields denotes the super-expansion of the superfields obtained after 
the application of HC (\ref{3.3}). Thus, from the above expressions, we can easily identify the (anti-)BRST symmetry 
transformations corresponding to the gauge field $A_\mu$ and (anti-)ghost fields $(\bar C)C$. These transformations 
are explicitly listed below
\begin{eqnarray}
&& s_b A_\mu = D_\mu C, \qquad s_b C =  \frac{g}{2}\, \big(C \times C\big), 
\qquad s_b \bar B = -\, g\,\big(\bar B \times C\big),\nonumber\\
&& s_b \bar C = i\, B,\qquad s_b B = 0, \label{3.6}
\end{eqnarray}  
\begin{eqnarray}
&&s_{ab} A_\mu = D_\mu \bar C, \qquad s_{ab} \bar C = \frac {g}{2} \,\big(\bar C \times \bar C\big), 
 \qquad s_{ab} B = - g\,\big(B \times \bar C\big), \nonumber\\
&& s_{ab} C = i \,\bar B,\qquad s_{ab} \bar B = 0. \label{3.7}
\end{eqnarray}
We point out that, the (anti-)BRST symmetry transformations for the Nakanishi-Lautrup auxiliary fields $B$ and $\bar B$ have been 
derived with the help of absolute anticommutativity and nilpotency properties of the above (anti-)BRST symmetries.

\subsection{(Anti-)BRST symmetries for $\phi_\mu$, $\beta$ and $\bar \beta$} 
In the previous subsection, we applied BT superfield approach to derive the off-shell nilpotent and absolutely anti-commuting 
(anti-)BRST symmetry transformations for the gauge field $(A_\mu)$ and corresponding (anti-)ghost fields $(\bar C)C$. Now, in 
order to derive the proper (anti-)BRST symmetries for the vector field $(\phi_\mu)$, corresponding (anti-)ghost fields 
$[(\bar \beta) \beta] $ and auxiliary field $(\rho)$, we have to go beyond the BT approach. For this purpose, we have 
exploited the power and strength of augmented superfield approach.

To derive the (anti-)BRST symmetries for the vector field $(\phi_\mu)$ and corresponding (anti-) ghost fields 
$[(\bar \beta) \beta]$, we invoke the following HC
\begin{eqnarray}
\tilde{\cal G}^{(2)} + \tilde {\mathscr{F}}^{(2)} \equiv G^{(2)} + {\mathscr{F}}^{(2)}, \label{hc}
\end{eqnarray}
where $G^{(2)}$, ${\mathscr{F}}^{(2)}$ are define in the following fashion
\begin{eqnarray}
 G^{(2)} &=&d\phi^{(1)} + i g\big(A^{(1)} \wedge \phi^{(1)}\big) + ig \big(\phi^{(1)} \wedge A^{(1)}\big)
= \frac{1}{2!}\,\big(dx^\mu \wedge dx^\nu\big)\,G_{\mu\nu},\nonumber\\
\mathscr{F}^{(2)} &=& -ig\big(F^{(2)} \wedge \rho^{(0)}\big) + ig\big(\rho^{(0)} \wedge F^{(2)}\big) 
= \frac{g}{2!}\,\big(dx^\mu \wedge dx^\nu\big)(F_{\mu\nu} \times \rho),
\end{eqnarray}
and $\tilde{\cal G}^{(2)}$,  $\tilde {\mathscr{F}}^{(2)}$ are the generalizations of $G^{(2)}$, ${\mathscr{F}}^{(2)}$ 
onto the superspace, respectively, which can be explicitly represented in the following manner 
\begin{eqnarray}
\tilde{\cal G}^{(2)}  &=& \tilde d \tilde \Phi^{(1)} 
+ i\, g\, \big(\tilde {\cal A}^{(1)}_{(h)} \wedge \tilde \Phi^{(1)}\big) 
+ i\, g\, \big(\tilde \Phi^{(1)} \wedge \tilde {\cal A}^{(1)}_{(h)}\big), \nonumber\\
\tilde {\mathscr{F}}^{(2)} &=& - i\, g\, \big(\tilde {\cal F}^{(2)}_{(h)} \wedge \tilde \rho^{(0)} \big) 
+ i\, g\, \big(\tilde \rho^{(0)} \wedge \tilde {\cal F}^{(2)}_{(h)}\big).
\end{eqnarray}
In the above expression, the quantities $\tilde {\cal A}^{(1)}_{(h)}, \tilde \Phi^{(1)}$ and $\tilde \rho^{(0)}$ are given as
\begin{eqnarray}
\tilde {\cal A}^{(1)}_{(h)} (x, \theta, \bar \theta) &=& dx^\mu \,\tilde {\cal A}^{(h)}_\mu(x, \theta, \bar \theta)
+ d\theta \,{\tilde {\bar {\cal F}}}^{(h)}(x, \theta, \bar \theta) 
+  d \bar \theta\, \tilde {\cal F}^{(h)}(x, \theta, \bar \theta), \nonumber\\
\tilde \Phi^{(1)} (x, \theta, \bar \theta) &=& dx^\mu\, \tilde\Phi_\mu(x, \theta, \bar\theta) 
+ d \theta \; \tilde {\bar \beta}(x, \theta, \bar\theta) 
+ d \bar \theta \; \tilde\beta(x, \theta, \bar\theta),\nonumber\\
\tilde \rho^{(0)} (x, \theta, \bar \theta) &=& \tilde \rho(x, \theta, \bar\theta), 
\end{eqnarray}
where the sub/super script $(h)$ denotes the quantities obtained after the application of HC. The superfields in the above 
expression, corresponding to the basic fields $\phi_\mu, \beta, \bar\beta$ and $\rho$ of the theory, can be expanded 
in terms of the secondary fields, as follows
\begin{eqnarray}
\tilde \Phi_\mu(x, \theta, \bar\theta) &=& \phi_\mu(x) + \theta\, \bar P_\mu(x) + \bar \theta\, P_\mu(x) 
+ i\,\theta\,\bar\theta\, Q_\mu(x),\nonumber\\
\tilde \beta(x, \theta, \bar\theta) &=& \beta(x) + i\, \theta\, \bar R_1(x) + i\, \bar \theta\, R_1(x) 
+ i\,\theta\,\bar\theta\, s_1(x),\nonumber\\
\tilde{\bar \beta}(x, \theta, \bar\theta) &=& \bar \beta(x) + i\, \theta\, \bar R_2(x) + i\, \bar \theta\, R_2(x) 
+ i\,\theta\,\bar\theta\, s_2(x),\nonumber\\
\tilde \rho (x, \theta, \bar\theta) &=&  \rho(x) + \theta\, \bar b(x) + \bar \theta\, b(x) 
+ i\,\theta\,\bar\theta\, q(x), \label{4.5}
\end{eqnarray}
where $P_\mu, \bar P_\mu, b, \bar b, s_1, s_2$ are fermionic secondary fields and 
$R_1, \bar R_1, R_2, \bar R_2, Q_\mu, q$ are bosonic in nature.

Exploiting the above HC (\ref{hc}) which demands that the coefficients of wedge products 
$(dx^\mu \wedge d \theta), \, (dx^\mu \wedge d \bar\theta),\, (d \theta \wedge d \theta),\,
(d \bar \theta \wedge d \bar \theta),\, (d \theta \wedge d \bar \theta)$ set equal to zero. We get following expressions:
\begin{eqnarray}
&&  \tilde {\cal D}_\mu \tilde {\bar \beta} - \partial_\theta \tilde \Phi_\mu  
- g\,\Big(\tilde \Phi_\mu \times {\tilde {\bar {\cal F}}}^{(h)}\Big) = 0, \qquad 
\partial_\theta \tilde {\bar \beta} - g\,\Big({\tilde {\bar {\cal F}}}^{(h)} \times \tilde  {\bar \beta}\Big) = 0,\nonumber\\
&&\tilde {\cal D}_\mu \tilde \beta - \partial_{\bar \theta} \tilde \Phi_\mu  
- g\,\Big(\tilde \Phi_\mu \times {\tilde {\cal  F}}^{(h)}\Big) = 0,\qquad 
\partial_{\bar \theta} \tilde \beta - g\,\Big({\tilde {\cal F}}^{(h)} \times \tilde \beta\Big) = 0,\nonumber\\ 
&& \partial_\theta \tilde \beta + \partial_{\bar \theta} \tilde {\bar \beta} 
- g\,\Big({\tilde {\bar {\cal F}}}^{(h)} \times \tilde  \beta\Big) 
- g\,\Big(\tilde{\cal  F}^{(h)} \times \tilde  {\bar \beta}\Big) =0, \label{4.6}
\end{eqnarray}
where $\tilde {\cal D}_\mu \bullet = \partial_\mu \bullet - g\big(\tilde {\cal A}^{(h)}_\mu \times \bullet \big)$. 
Using the expansion (\ref{4.5}) in (\ref{4.6}),
we get following relationships amongst the basic and secondary fields of the theory, namely;
\begin{eqnarray}
R_1 &=& - i\,g (C \times \beta), \quad \bar R_2 = - i\,g (\bar C \times \bar \beta), \quad 
s_1 = - g (\bar B \times \beta) + g (C \times \bar R),\nonumber\\
s_2 &=& g\,(B \times \bar \beta) - g\,(\bar C \times R), 
\quad R + \bar R + i\, g(C \times \bar \beta) + i \,g (\bar C \times \beta)= 0, \nonumber\\
 P_\mu &=& D_\mu \beta - g (\phi_\mu \times C), \quad 
D_\mu \bar R_2 + i\, g  (D_\mu \bar C \times \bar \beta) + i\, g (D_\mu \bar \beta \times \bar C) = 0, \nonumber\\
 \bar P_\mu &=& D_\mu \bar \beta - g (\phi_\mu \times  \bar C), 
\quad D_\mu R_1 + i \,g (D_\mu  C \times \beta) + i \,g (D_\mu \beta \times C) = 0,\nonumber\\
 Q_\mu &=& D_\mu R + g (B \times \phi_\mu) + i\, g(D_\mu C \times \bar \beta) 
+ i \,g[D_\mu \beta \times \bar C - g(\phi_\mu \times C)\times \bar C]\nonumber\\
 &\equiv& - D_\mu \bar R - g (\bar B \times \phi_\mu) - i\, g(D_\mu \bar C \times  \beta) 
- i \,g[D_\mu \bar \beta \times C - g(\phi_\mu \times \bar C)\times C], \nonumber\\ \label{3.14}
\end{eqnarray}
where we have chosen $\bar R_1 = \bar R$, $R_2 = R$.
Substituting, these values of secondary fields in (\ref{4.5}), we have following form of superfield expansions
\begin{eqnarray}
 {\tilde \Phi_\mu}^{(h)} (x, \theta, \bar \theta) &=& \phi_\mu(x) + \theta \big[D_\mu \bar \beta - g\,(\phi_\mu \times \bar C)\big](x)
+ \bar  \theta \big[D_\mu \beta - g\,(\phi_\mu \times  C)\big](x)\nonumber\\
 &+& \theta\, \bar \theta\,\big[i \,D_\mu R - i\, g\, (\phi_\mu \times B) - g\,(D_\mu C \times \bar \beta) -  g\,(D_\mu \beta \times \bar C) \nonumber\\ 
 &+& g^2\,(\phi_\mu \times C)\times \bar C \big](x) \nonumber\\
  &\equiv& \phi(x) + \theta \,\big(s_b \,\phi(x)\big) + \bar \theta \,\big(s_{ab} \,\phi (x)\big)
 + \theta\,\bar\theta \,\big(s_b \,s_{ab}\,\phi(x)\big),\nonumber\\
  {\tilde \beta}^{(h)} (x, \theta, \bar \theta) &=& \beta (x) + \theta \,\big[i\, \bar R (x)\big]
+ \bar  \theta\, \big[g\, (C \times \beta)\big](x) \nonumber\\
 &+& \theta\, \bar \theta\,\big[- i\, g\,(\bar B \times \beta) - i\, g\, (\bar R \times C) \big](x) \nonumber\\ 
 &\equiv& \beta(x) + \theta \,\big(s_b \,\beta(x)\big) + \bar \theta \,\big(s_{ab} \,\beta (x)\big)
 + \theta\,\bar\theta \,\big(s_b \,s_{ab}\,\beta(x)\big),\nonumber\\
 {\tilde {\bar \beta}}^{(h)} (x, \theta, \bar \theta) &=& \bar \beta (x) + \theta\, \big[g \,(\bar C \times \bar \beta)\big](x)
+ \bar  \theta \,\big[i\, R(x)\big] \nonumber\\
  &+& \theta \,\bar \theta\,\big[i \,g \,(B \times \bar \beta) + i\, g \,(R \times \bar C)\big](x) \nonumber\\ 
  &\equiv& \bar \beta(x) + \theta \,\big(s_b \,\bar \beta(x)\big) + \bar \theta \,\big(s_{ab} \,\bar \beta (x)\big)
 + \theta\,\bar\theta \,\big(s_b \,s_{ab}\,\bar \beta(x)\big), \label{3.15}
\end{eqnarray}
here $(h)$ on the superscript of superfields represents the respective quantities obtained after the application of HC (\ref{hc}). Therefore, 
(anti-)BRST symmetry transformations for vector field $(\phi_\mu)$ and (anti-)ghost fields $[(\bar \beta)\beta]$ are obvious 
from the above super-expansions.

\subsection{(Anti-)BRST symmetries for auxiliary field $\rho$}
In order to derive the proper (anti-)BRST symmetry transformations for the auxiliary field $\rho$, we look for a quantity 
which remains invariant (or should transform covariantly) under the combined gauge transformations (\ref{2.5}). Such gauge 
invariant quantity will serve a purpose of `physical quantity' (in some sense) which could be generalized onto the $(3,2)$-dimensional 
supermanifold. Furthermore, being a `physical quantity' it should remain unaffected by the presence of Grassmannian 
variables when the former is generalized onto the supermanifold. Thus, keeping above in mind, we note that under the 
combined gauge transformations (\ref{2.5}), the  quantity $(D_\mu \rho - \phi_\mu)$ transforms covariantly (as the 
quantities $F_{\mu\nu}$ and  $G_{\mu\nu} + g (F_{\mu\nu} \times \rho)$ do). This can be explicitly checked as follows
\begin{eqnarray}
\delta(D_\mu \rho - \phi_\mu) = - g \,(D_\mu \rho - \phi_\mu)\times \Lambda.
\end{eqnarray}
Therefore, the above quantity serves our purpose and it can also be expressed in the language of differential forms as follows
\begin{eqnarray}
d \rho^{(0)} + i\, g\, \big(A^{(1)} \wedge \rho^{(0)}\big) -  i\, g\, \big(\rho^{(0)} \wedge A^{(1)}\big) - \phi^{(1)}
 &=& dx^\mu \big(D_\mu \rho - \phi_\mu \big), 
\end{eqnarray}
which is clearly a 1-form object. Now, we generalize this 1-form object onto the $(3, 2)$-dimensional supermanifold 
and demand that it should remain unaffected by the presence of Grassmannian variables. This, in turn, produces the following HC 
\begin{eqnarray}
 d \rho^{(0)} + i\, g \big(A^{(1)} \wedge \rho^{(0)}\big) -  i\, g \big(\rho^{(0)} \wedge A^{(1)}\big) - \phi^{(1)} &\equiv& 
\tilde d \tilde \rho^{(0)} + i\, g \big(\tilde {\cal A}^{(1)}_{(h)} \wedge \tilde \rho^{(0)}\big) \nonumber\\
&-&  i \,g \big(\tilde \rho^{(0)} \wedge \tilde {\cal A}^{(1)}_{(h)}\big) - \tilde \Phi^{(1)}_{(h)}. \qquad\label{gir}
\end{eqnarray}
This HC can also be derived from the integrability of (\ref{hc}) (see e.g., \cite{ThierryMieg:1982un} 
for details on the topic). The r.h.s. of the above HC can be simplified as
\begin{eqnarray}
&& \tilde d \tilde \rho^{(0)} + i g \big(\tilde {\cal A}^{(1)}_{(h)} \wedge \tilde \rho^{(0)}\big) 
-  i g \big(\tilde \rho^{(0)} \wedge \tilde {\cal A}^{(1)}_{(h)} \big) - \tilde \Phi^{(1)}_{(h)} = \nonumber\\  
&& dx^\mu\Big[{\tilde {\cal D}_\mu} \tilde \rho - \tilde \Phi_\mu^{(h)}\Big] 
+ d\theta \Big[\partial_\theta \tilde \rho - \tilde{\bar \beta}^{(h)} 
- g \Big({\tilde {\bar {\cal F}}}^{(h)} \times \tilde \rho\Big)\Big]
+ d\bar \theta \Big[\partial_{\bar \theta} \tilde \rho - \tilde \beta^{(h)}
- g \Big(\tilde {\cal F}^{(h)} \times \tilde \rho\Big)\Big]. \qquad \;
\end{eqnarray}
Exploiting (\ref{gir}), and set the coefficients of $d\theta, d\bar \theta$ equal to zero, we have the following relationships, namely;
\begin{eqnarray}
\partial_\theta \tilde \rho - \tilde{\bar \beta}^{(h)} - g\,\Big({\tilde {\bar {\cal F}}}^{(h)} \times \tilde \rho\Big) =0,\qquad 
\partial_{\bar \theta} \tilde \rho - \tilde \beta^{(h)} - g\,\Big(\tilde {\cal F}^{(h)} \times \tilde \rho\Big) =0.
\end{eqnarray}
Plugging the values of superfield expansions from (\ref{3.5}), (\ref{4.5}) and (\ref{3.15}) into the above expressions, we get 
the following relationships amongst
the basic and secondary fields
\begin{eqnarray}
b &=& \beta - g(\rho \times  C), \qquad \bar b = \bar \beta - g(\rho \times \bar C),\nonumber\\ 
q &=& R +  g(B \times \rho) + i g (\bar C \times \beta) - i g^2(\rho \times C)\times \bar C\nonumber\\
&\equiv& - \bar R - i g (C \times \bar \beta) - g (\bar B \times \rho) + i g^2 (\rho \times \bar C)\times C.
\end{eqnarray}
We point out that, however, there also exist other relationships but they are same as quoted in equation (\ref{3.14}).

Finally, substituting these values of secondary fields into (\ref{4.5}), we obtain the  following superfield expansion 
for the super-auxiliary field $\tilde \rho (x, \theta, \bar\theta)$
\begin{eqnarray}
 \tilde \rho^{(h)} (x, \theta, \bar \theta) &=& \rho (x) + \theta \,\big[\bar \beta - g\,(\rho \times \bar C)\big](x)
+ \bar  \theta \,\big[\beta - g\,(\rho \times  C)\big](x) \nonumber\\
&+ & i \theta\, \bar \theta\, \big[R +  g\,(B \times \rho) + i\, g\, (\bar C \times \beta) 
- i \,g^2\,(\rho \times C)\times \bar C\big](x), \nonumber\\
&\equiv& \rho(x) + \theta \,\big(s_b \,\rho(x)\big) + \bar \theta \,\big(s_{ab} \,\rho (x)\big)
 + \theta\,\bar\theta \,\big(s_b \,s_{ab}\,\rho(x)\big), \label{3.22}
\end{eqnarray}
where $(h)$ as the superscript on the generic superfield denotes the corresponding superfield expansion obtained 
after the application of HC (\ref{gir}). The (anti-) BRST symmetry transformations for the auxiliary field $\rho$ 
can be easily deduced from the above expansion. Thus, we have derived the proper (anti-)BRST symmetry transformations 
for the vector field $(\phi_\mu)$, corresponding (anti-)ghost fields $[(\bar \beta)\beta]$ and auxiliary field 
$(\rho)$ within the framework of augmented superfield formalism. Moreover, the (anti-)BRST symmetry transformations for the Nakanishi-Lautrup
auxiliary fields $R$ and $\bar R$ have been derived with the help of anticommutativity and nilpotency properties of the (anti-)BRST symmetries. 
These symmetry transformations are listed below
\begin{eqnarray}
&& s_b \phi_\mu = D_\mu \beta - g \big(\phi_\mu \times C\big), \quad s_b \beta = g \big(C \times \beta\big), 
\quad s_b \rho = \beta -\, g \big(\rho \times C\big), \nonumber\\
&& s_b \bar \beta = i\, R, \quad  s_b R = 0,  \quad s_b \bar R = - g \big(\bar R \times C\big) 
- g \big(\bar B \times \beta\big), \label{3.23}
\end{eqnarray}
\begin{eqnarray}
&&s_{ab} \phi_\mu = D_\mu \bar \beta - g \big(\phi_\mu \times \bar C\big), 
\quad s_{ab} \bar \beta = g \big(\bar C \times \bar \beta\big), 
\quad  s_{ab} \rho = \bar \beta - g \big(\rho \times \bar C\big),\nonumber\\
&&s_{ab} \beta = i \,\bar R, \quad s_{ab} \bar R = 0, \quad  
\quad s_{ab} R = - g \big(R \times \bar C\big) - g \big(B \times \bar \beta\big). \label{3.24}
\end{eqnarray}
These (anti-)BRST symmetry transformations as well as the transformations listed in (\ref{3.6}) and (\ref{3.7}) 
are off-shell nilpotent $(s_{(a)b}^2 \Psi = 0)$ and absolutely anticommuting 
$[(s_b s_{ab} + s_{ab} s_b) \Psi = 0]$ in nature. Here $\Psi$ represents any generic field of the theory. 
These properties (i.e. nilpotency and anticommutativity) are two key ingredients of the BRST formalism. 
The anticommutativity property for the vector fields ($\phi_\mu$ and $A_\mu$) and 
auxiliary field ($\rho$) is satisfied only on the constrained surface parametrized by the 
CF conditions (cf. (\ref{3.26}) below). For instance, one can check that
\begin{eqnarray}
&&\{s_b,\, s_{ab}\} A_\mu = iD_\mu \big[B + \bar B + i (C \times \bar C)],\nonumber\\
&& \{s_b, \,s_{ab}\} \phi_\mu = iD_\mu\big[R + \bar R + i g(C \times \bar \beta) + i g (\bar C \times \beta) \big]
+ i g \big[B + \bar B + i g (C \times \bar C)\big] \times \phi_\mu, \nonumber\\
&& \{s_b, \,s_{ab}\} \rho = i\big[R + \bar R + i g(C \times \bar \beta) + i g (\bar C \times \beta) \big]
+ i g \big[B + \bar B + i g (C \times \bar C)\big] \times \rho, 
\end{eqnarray} 
whereas, for all the {\it rest} of the fields (of our present 3D JP model), the absolute anticommutativity property 
(i.e. $ \{s_b,  s_{ab} \} \Psi = 0$) is valid {\it without} invoking the CF type conditions.

Before, we wrap up this section, some crucial points are in order. First and foremost, a very careful look at 
(\ref{3.4}) and (\ref{3.14}) reveals, respectively, the existence of two sets of Curci-Ferrari (CF) type conditions, namely; 
\begin{eqnarray}
&(i)& B + \bar B + i g (C \times \bar C) = 0, \nonumber\\
&(ii)& R + \bar R + i \,g\,\big(C \times \bar \beta\big) + i\, g\,\big(\bar C \times \beta\big) = 0. \label{3.26}
\end{eqnarray}
These conditions are key signatures of any $p$-form gauge theory when the latter is discussed 
within the framework of BRST formalism. In our case, the above mentioned CF conditions emerge very naturally 
within the framework of (augmented) superfield formalism. 
In fact, CF conditions $(i)$ and $(ii)$  emerge from the HC (\ref{3.3}) and (\ref{hc}), respectively, when we set the   
coefficients of $(d\theta \wedge d \bar \theta)$ equal to zero.
Second, the absolute anticommutativity of (anti-)BRST 
symmetries is ensured by these CF type conditions. Third, these CF type conditions are (anti-)BRST invariant.
Finally, these CF type conditions play a crucial role in the derivation of the coupled (but equivalent) Lagrangian 
densities. We have discussed this aspect, in detail, in our next section.

\section{Coupled Lagrangian densities}
In this section, we construct the coupled (but equivalent) Lagrangian densities which respect nilpotent as well as 
anticommuting (anti-)BRST symmetry transformations derived in the previous section (cf. Section 3). In order to proceed 
further, a few important points are in order. First, the mass dimensions (in natural units $c = \hbar =1$) of the 
various fields in our present 3D theory are: $[A_\mu] = [\phi_\mu] = [C] = [\bar C] = [\beta] = [\beta] 
= [M]^{\frac{1}{2}}, \; [B] = [\bar B] = [R] = [\bar R] = [M]^{\frac{3}{2}},\;
[\rho] = [M]^{-\frac{1}{2}}, \;$ and the coupling constant $g$ has the mass dimension $[g] = [M]^{\frac{1}{2}}$.
Second, the fermionic (anti-)ghost fields $(\bar C) C$ and $(\bar \beta) \beta$ carry ghost numbers $(\mp1)$, 
respectively whereas  rest of the (bosonic) fields carry ghost number equal to zero. Third, the nilpotent (anti-)BRST transformations 
increase the mass dimension by one unit when they operate on any generic field of the theory. In other words, we can say that the
(anti-)BRST transformations carry mass dimension equal one (in natural units). Fourth, the (anti-)BRST transformations 
(decrease)increase the ghost number by one unit when they act on any field of the theory. This means that (anti-)BRST transformations 
carry ghost number $(\mp 1)$, respectively. These points are very important in constructing the (anti-)BRST invariant
coupled Lagrangian densities.

Exploiting the basic tenets of the BRST formalism, the most appropriate (anti-)BRST invariant Lagrangian densities that can be 
written in terms of nilpotent and absolutely anticommuting (anti-)BRST symmetry transformations are as follows \cite{ThierryMieg:1982un}
\begin{eqnarray}
{\cal L}_b &=& {\cal L}_0 + s_b\,s_{ab}\bigg[\frac{i}{2} A_\mu \cdot A^\mu + C \cdot \bar C 
+ \frac{i}{2} \phi_\mu \cdot \phi^\mu + \frac{1}{2}\, \beta \cdot \bar \beta\bigg],\nonumber\\
&&\nonumber\\
{\cal L}_{\bar b} &=& {\cal L}_0 - s_{ab}\,s_b\bigg[\frac{i}{2} A_\mu \cdot A^\mu + C \cdot \bar C 
+ \frac{i}{2} \phi_\mu \cdot \phi^\mu + \frac{1}{2}\,\beta \cdot \bar \beta\bigg],
\end{eqnarray}
where ${\cal L}_0$ is our starting gauge invariant Lagrangian density (\ref{2.1}). We would like to emphasize that each term 
in the square brackets is Lorentz scalar and chosen in such a way that they have ghost number zero and 
mass dimension one (in natural units). Moreover, the (constant) factors in front
of each term are picked for the algebraic convenience.  Utilizing the off-shell nilpotent (anti-) BRST transformations from 
(\ref{3.6}), (\ref{3.7}), (\ref{3.23}) and (\ref{3.24}), we obtain the following explicit  Lagrangian densities, namely;
\begin{eqnarray}
{\cal L}_b &=& - \frac{1}{4}\, F_{\mu\nu}\cdot F^{\mu\nu} 
- \frac{1}{4}\, \big(G_{\mu\nu} + g F_{\mu\nu} \times \rho\big) \cdot \big(G^{\mu\nu} + g F^{\mu\nu} \times \rho\big) 
+ \frac{m}{2}\, \varepsilon^{\mu\nu\eta}\,F_{\mu\nu}\cdot \phi_\eta\nonumber\\
&+& \frac{1}{2}\,\big[B \cdot B + \bar B \cdot \bar B\big] + B\cdot \big(\partial^\mu A_\mu\big)
+ \frac{1}{2}\,\big[R + i g (C \times \bar \beta)\big]\cdot \big[R + i g (C \times \bar \beta)\big]\nonumber\\
&+& \big[R + i g (C \times \bar \beta)\big]\cdot \big(D^\mu \phi_\mu\big) 
- i \partial_\mu \bar C \cdot D^\mu C - i D_\mu \bar \beta \cdot D^\mu \beta,\nonumber\\
&&\nonumber\\
{\cal L}_{\bar b} &=& - \frac{1}{4}\, F_{\mu\nu}\cdot F^{\mu\nu} 
- \frac{1}{4}\, \big(G_{\mu\nu} + g F_{\mu\nu} \times \rho\big) \cdot \big(G^{\mu\nu} + g F^{\mu\nu} \times \rho\big) 
+ \frac{m}{2}\, \varepsilon^{\mu\nu\eta}\,F_{\mu\nu}\cdot \phi_\eta\nonumber\\
&+& \frac{1}{2}\,\big[B \cdot B + \bar B \cdot \bar B\big] - \bar B\cdot \big(\partial^\mu A_\mu\big)
+ \frac{1}{2}\,\big[\bar R + i g (\bar C \times \beta)\big]\cdot \big[\bar R + i g (\bar C \times \beta)\big]\nonumber\\
&-& \big[\bar R + i g (\bar C \times \beta)\big]\cdot \big(D^\mu \phi_\mu\big) 
- i D_\mu \bar C \cdot \partial^\mu C - i D_\mu \bar \beta \cdot D^\mu \beta, \label{4.2}
\end{eqnarray}
where $B, \bar B$ and $R, \bar R$ are the Nakanishi-Lautrup type auxiliary fields. These Lagrangian densities 
are coupled because these Nakanishi-Lautrup auxiliary fields $B, \bar B$ and $R, \bar R$  are related through the 
CF conditions (\ref{3.26}).

It can be checked that the (anti-)BRST transformations ($s_{(a)b}$) leave the above Lagrangian densities 
quasi-invariant. To be more specific,  under  the operations of nilpotent (anti-)BRST transformations, the  Lagrangian 
densities $({\cal L}_{\bar b}) {\cal L}_b$ transform to a total spacetime derivative, in the following fashion, respectively  
\begin{eqnarray}
s_{ab} {\cal L}_{\bar b} &=& \partial_\mu \bigg[\frac{m}{2}\, \varepsilon^{\mu\nu\eta} F_{\nu\eta} \cdot \bar \beta 
- \bar B \cdot (D^\mu \bar C) - \bar R\cdot D^\mu \bar \beta - i g \big(\bar C \times \beta\big)\cdot D^\mu \bar \beta\bigg], \nonumber\\
&&\nonumber\\
s_b {\cal L}_b &=& \partial_\mu \bigg[\frac{m}{2}\, \varepsilon^{\mu\nu\eta} F_{\nu\eta} \cdot \beta + B \cdot (D^\mu C)
+ R\cdot D^\mu \beta + i g \big(C \times \bar \beta\big)\cdot D^\mu \beta\bigg]. \quad
\end{eqnarray}
Thus,  the action integral corresponding to the above Lagrangian densities remain invariant 
under ($s_{(a)b}$). Furthermore, it is interesting to note that the following variations are true:
\begin{eqnarray}
s_{ab} {\cal L}_b &=& \partial_\mu \Big[\frac{m}{2}\, \varepsilon^{\mu\nu\eta} F_{\nu\eta} \cdot \bar \beta 
+ B \cdot \partial^\mu \bar C + \big(R + i g C \times \bar \beta \big)\cdot D^\mu \bar \beta  \Big]\nonumber\\
&-& \Big[D_\mu\big(B + \bar B + i g C \times \bar C\big)\Big]\cdot \partial^\mu \bar C 
- \Big[D_\mu\big(R + \bar R + ig C \times \bar \beta + i g \bar C \times \beta \big)\Big] \cdot D^\mu\bar \beta\nonumber\\
&-& g \Big[R + ig \big(C \times \bar \beta\big) 
+ D^\mu \phi_\mu \Big]\cdot \Big[\big(B + \bar B + i g C \times \bar C\big)\times \bar \beta\Big], \nonumber\\
&&\nonumber\\
s_b {\cal L}_{\bar b} &=& \partial_\mu \Big[\frac{m}{2}\, \varepsilon^{\mu\nu\eta} F_{\nu\eta} \cdot  \beta 
- \bar B \cdot \partial^\mu  C - \big(\bar R + i g \bar C \times  \beta \big)\cdot D^\mu  \beta \Big]\nonumber\\
&+& \Big[D_\mu\big(B + \bar B + i g C \times \bar C\big)\Big]\cdot \partial^\mu C 
+ \Big[D_\mu\big(R + \bar R + ig C \times \bar \beta + i g \bar C \times \beta \big)\Big] \cdot D^\mu \beta \nonumber\\
&-& g \big[\bar R + ig \big(\bar C \times \beta \big) 
- D^\mu \phi_\mu\Big]\cdot \Big[\big(B + \bar B + i g C \times \bar C\big)\times  \beta\Big].
\end{eqnarray}
Therefore, it is evident from the above variations that the Lagrangian densities  ${\cal L}_b$ and ${\cal L}_{\bar b}$
also respect the anti-BRST ($s_{ab}$) and BRST ($s_b$) transformations, respectively only on the constrained hypersurface defined 
by the CF conditions (\ref{3.26}). As a result, both the Lagrangian densities are equivalent and they respect BRST as well as anti-BRST symmetries 
on the constrained hypersurface spanned by CF conditions [cf. (\ref{3.26})].

\section{Conserved charges: Novel observations}
In our previous section, we have seen that the coupled Lagrangian densities (and corresponding actions) respect
the off-shell nilpotent and continuous (anti-)BRST symmetry transformations. As a consequence, according to Noether's theorem,
the invariance of the actions under the continuous (anti-) BRST transformations lead to the following conserved 
(anti-)BRST currents ($J^\mu_{(a)b}$), namely;  
\begin{eqnarray}
J^\mu_{ab} &=& - (D_\nu \bar C) \cdot \Big[F^{\mu\nu} - g \big(G^{\mu\nu} + g F^{\mu\nu} \times \rho\big)\times \rho 
- m \,\varepsilon^{\mu\nu\eta} \phi_\eta\Big] - \bar B \cdot (D^\mu \bar C) \nonumber\\
&-& \frac{i}{2}\,g \big(\bar C \times \bar C\big) \cdot \partial^\mu  C 
- (D_\nu \bar \beta)\cdot \Big(G^{\mu\nu} + g F^{\mu\nu} \times \rho \Big) 
+ g (\phi_\nu \times \bar C) \cdot \Big(G^{\mu\nu} + g F^{\mu\nu} \times \rho \Big) \nonumber\\
&+&  g (\phi^\mu \times \bar C)\cdot \big(\bar R + i g \bar C\times \beta\big) 
- \bar R\cdot D^\mu \bar \beta - i g \big(\bar C \times \bar \beta \big)\cdot D^\mu \beta  
- \frac{m}{2}\,\varepsilon^{\mu\nu\eta} F_{\nu\eta} \cdot \bar  \beta,\nonumber\\
&&\nonumber\\ 
J^\mu_b &=& - (D_\nu C) \cdot \Big[F^{\mu\nu} - g \big(G^{\mu\nu} + g F^{\mu\nu} \times \rho\big)\times \rho 
- m \,\varepsilon^{\mu\nu\eta} \phi_\eta\Big] + B \cdot (D^\mu C)\nonumber\\
&+& \frac{i}{2}\,g \big(C \times C\big) \cdot \partial^\mu \bar C - (D_\nu \beta)\cdot \Big(G^{\mu\nu} + g F^{\mu\nu} \times \rho \Big) 
+ g (\phi_\nu \times C) \cdot \Big(G^{\mu\nu} + g F^{\mu\nu} \times \rho \Big)\nonumber\\ 
&-&  g (\phi^\mu \times C)\cdot \big(R + i g C\times \bar \beta\big) 
+ R\cdot D^\mu \beta + i g \big(C \times \beta \big)\cdot D^\mu \bar \beta  
- \frac{m}{2}\,\varepsilon^{\mu\nu\eta} F_{\nu\eta} \cdot \beta.
\end{eqnarray}
One can check that the conservation  (i.e. $\partial_\mu J^\mu_b = 0$) of BRST current ($J^\mu_b$) can be proven by 
exploiting the Euler-Lagrange (E-L) equations of motion that are derived from the  
Lagrangian density ${\cal L}_b$. These E-L equations are as listed below:
\begin{eqnarray}
&& D_\mu F^{\mu\nu} - g\,D_\mu\big[\big(G^{\mu\nu} + g\, F^{\mu\nu} \times \rho \big)\times \rho\big] 
+ g \,\big(G^{\mu\nu} + g\, F^{\mu\nu} \times \rho \big) \times \phi_\mu - m \,\varepsilon^{\mu\nu\eta}\,(D_\mu \phi_\eta)
 \nonumber\\
&& - \; \partial^\nu  B - i g \big(\partial^\nu \bar C \times C\big)
+ g \big(R + i g C \times \bar \beta\big)\times \phi^\nu 
+ i\, g \big(\bar \beta \times D^\nu \beta\big) - i\,g \big(\beta \times D^\nu \bar \beta\big) = 0, \nonumber\\
&& D_\mu \big[G^{\mu\nu} + g (F^{\mu\nu} \times \rho) \big] - D^\nu \big[R + i g(C \times \bar \beta)\big] 
- \frac{m}{2}\, \varepsilon^{\mu\nu\kappa}\, F_{\mu\kappa} = 0, \nonumber\\
&& \big[G^{\mu\nu} + g (F^{\mu\nu} \times \rho) \big] \times F_{\mu\nu} = 0, \qquad R + ig (C \times \bar \beta) + D_\mu \phi^\mu = 0,
 \qquad B = - \big(\partial_\mu A^\mu\big), \nonumber\\
 && \partial_\mu (D^\mu C) = 0, \qquad D_\mu (\partial^\mu \bar C) = 0, \qquad D_\mu (D^\mu \beta) = 0. 
\qquad D_\mu (D^\mu \bar \beta) = 0,  \label{5.2}
\end{eqnarray}
\begin{eqnarray}
&& D_\mu F^{\mu\nu} - g\,D_\mu\big[\big(G^{\mu\nu} + g\, F^{\mu\nu} \times \rho \big)\times \rho\big] 
+ g \,\big(G^{\mu\nu} + g\, F^{\mu\nu} \times \rho \big) \times \phi_\mu - m \,\varepsilon^{\mu\nu\eta}\,(D_\mu \phi_\eta) 
 \nonumber\\
&& +\; \partial^\nu \bar B +  i g \big(\partial^\nu C \times \bar C\big)
- g \big(\bar R + i g \bar C \times \beta\big)\times \phi^\nu 
+ i\, g \big(\bar \beta \times D^\nu \beta\big) - i\,g \big(\beta \times D^\nu \bar \beta\big) = 0, \nonumber\\
&& D_\mu \big[G^{\mu\nu} + g (F^{\mu\nu} \times \rho) \big] + D^\nu \big[\bar R + i g(\bar C \times  \beta)\big] 
- \frac{m}{2}\, \varepsilon^{\mu\nu\kappa}\, F_{\mu\kappa} = 0, \nonumber\\
&& \big[G^{\mu\nu} + g (F^{\mu\nu} \times \rho) \big] \times F_{\mu\nu} = 0, \qquad \bar R + ig (\bar C \times \beta) - D_\mu \phi^\mu = 0,
 \qquad \bar B =  \big(\partial_\mu A^\mu\big), \nonumber\\
 && D_\mu (\partial^\mu C) = 0, \qquad \partial_\mu (D^\mu \bar C) = 0, \qquad D_\mu (D^\mu \beta) = 0, 
\qquad D_\mu (D^\mu \bar \beta) = 0,  \label{5.3}
\end{eqnarray}
which emerge  from the Lagrangian density ${\cal L}_{\bar b}$.

Exploiting the above E-L equations of motion (cf. (\ref{5.2}) and (\ref{5.3})),  the conserved currents 
$J^\mu_{(a)b}$ can be written in simpler forms as:
\begin{eqnarray}
J^\mu_{ab} &=& - \partial_\nu \Big(\big[F^{\mu\nu} - g \big(G^{\mu\nu} + g F^{\mu\nu} \times \rho\big)\times \rho 
- m \,\varepsilon^{\mu\nu\eta} \phi_\eta\big]\cdot \bar C + \big[G^{\mu\nu} + g F^{\mu\nu} \times \rho\big]\cdot \bar \beta\Big)\nonumber\\
&+& (\partial^\mu \bar B) \cdot \bar C  - \bar B \cdot (D^\mu \bar C) 
+ \frac{i}{2}\,g \big(\bar C \times \bar C\big) \cdot \partial^\mu  C 
- (\bar R + i g \bar C \times \beta \big)\cdot (D^\mu \bar \beta) \nonumber\\
&+& D^\mu\big(\bar R + i g \bar C \times \beta\big) \cdot \bar \beta,\nonumber\\
&&\nonumber\\
J^\mu_b &=& - \partial_\nu \Big(\big[F^{\mu\nu} - g \big(G^{\mu\nu} + g F^{\mu\nu} \times \rho\big)\times \rho 
- m \,\varepsilon^{\mu\nu\eta} \phi_\eta\big]\cdot C + \big[G^{\mu\nu} + g F^{\mu\nu} \times \rho\big]\cdot \beta\Big)\nonumber\\
&+& B \cdot (D^\mu C) - (\partial^\mu B) \cdot C - \frac{i}{2}\,g \big(C \times C\big) \cdot \partial^\mu \bar C 
+ (R + i g C \times \bar \beta \big)\cdot (D^\mu \beta)\nonumber\\
&-& D^\mu\big(R + i g C \times \bar \beta\big)\cdot \beta. 
\end{eqnarray}
Now, the proof of conservation laws ($\partial_\mu J^\mu_{(a)b} = 0$) is quite straightforward.  
The temporal components (i.e. $\int d^2x J^0_{(a)b} = Q_{(a)b}$) of the above conserved currents ($J^\mu_{(a)b}$)
 lead to the following conserved (i.e. ${\dot Q}_{(a)b} =0$) (anti-)BRST charges ($Q_{(a)b}$), namely; 
\begin{eqnarray}
Q_{ab} &=& - \int d^2x \bigg[\bar B \cdot (D^0 \bar C) - (\partial^0 \bar B) \cdot \bar C 
- \frac{i}{2}\,g \big(\bar C \times \bar C\big) \cdot \partial^0 C 
+ (\bar R + i g \bar C \times \beta \big)\cdot (D^0 \bar \beta)\nonumber\\ 
&-& D^0\big(\bar R + i g \bar C \times \beta\big)\cdot \bar \beta \bigg], \nonumber\\ 
&&\nonumber\\
Q_b &=& \int d^2x \bigg[ B \cdot (D^0 C) - (\partial^0 B) \cdot C 
- \frac{i}{2}\,g \big(C \times C\big) \cdot \partial^0 \bar C 
+ (R + i g C \times \bar \beta \big)\cdot (D^0 \beta) \nonumber\\
&-& D^0\big(R + i g C \times \bar \beta\big)\cdot \beta \bigg].\label{5.5}
\end{eqnarray}
It turns out that the conserved, nilpotent ($Q^2_{(a)b} = 0$, see below) and anticommuting
($Q_b\,Q_{ab} + Q_{ab}\, Q_b = 0$, see below) (anti-)BRST charges are the generators of the (anti-)BRST 
symmetry transformations, respectively. For the sake of brevity, these transformations can be obtained by 
exploiting the following symmetry properties:  
\begin{eqnarray}  
s_b \Psi  = - i \big[\Psi, \; Q_b\big]_{\pm}, \qquad s_{ab} \Psi  = - i \big[\Psi,\; Q_{ab} \big]_{\pm},\qquad
\Psi = A_\mu, \phi_\mu, C, \bar C, \beta, \bar \beta
\end{eqnarray}
The $(\pm)$ signs as the subscript on the square brackets represent (anti)commutators 
corresponding to the generic field $\Psi$ being (fermionic)bosonic in nature (see, e.g. \cite{Gupta:2009bu} for details). 
The (anti-)BRST transformations of the Nakanishi-Lautrup auxiliary fields 
$B, \bar B, R, \bar R$ have been derived from the basic requirements (i.e. nilpotency and/or absolute 
anticommutativity properties) of the (anti-)BRST symmetry transformations.

It is worthwhile to mention that, even though, the (anti-)BRST charges $(Q_{(a)b})$ are conserved, nilpotent 
as well as anticommuting in nature (see below), they are unable to generate the proper (anti-)BRST transformations
(i.e. $s_b \rho = \beta - g (\rho \times C)$ and $s_{ab} \rho = \bar \beta - g (\rho \times \bar C)$) 
of the auxiliary field $\rho$. Furthermore, the nilpotency and absolute anticommutativity
properties of the (anti-)BRST transformations also fail to produce the transformations of $\rho$. This is one of the 
novel observations of our present endeavor. Although, we have derived these transformations by exploiting the power and strength of
the augmented superfield formalism which produces the off-shell nilpotent ($s^2_{(a)b} = 0$) as well as absolutely anticommuting
($s_b \,s_{ab} + s_{ab}\, s_b = 0$) (anti-)BRST symmetry transformations for {\it all} the basic and auxiliary fields of the theory.

The nilpotency ($Q^2_{(a)b} =0$) of the (anti-)BRST charges reflects the fermionic nature whereas the anticommutativity 
($Q_b\,Q_{ab} + Q_{ab}\, Q_b = 0$) shows that the (anti-)BRST charges are linearly independent of each other. 
These properties can be verified in the following straightforward manner:
\begin{eqnarray}
s_b Q_b &=& - i \{Q_b,\; Q_b\} = 0 \Rightarrow Q^2_b = 0,\nonumber\\
s_{ab} Q_{ab} &=& - i \{Q_{ab},\; Q_{ab}\} = 0 \Rightarrow Q^2_{ab} = 0,\nonumber\\
s_b Q_{ab} &=& - i \{Q_{ab},\; Q_b\} = 0 \Rightarrow Q_{ab}\,Q_{b} + Q_{b}\, Q_{ab} = 0,\nonumber\\
s_{ab} Q_b &=& - i \{Q_b,\; Q_{ab}\} = 0 \Rightarrow Q_b\,Q_{ab} + Q_{ab}\, Q_b = 0.
\end{eqnarray}
We point out that in proving the anticommutativity property ($Q_b\,Q_{ab} + Q_{ab}\, Q_b = 0$) of the 
(anti-)BRST charges we have used the CF conditions (\ref{3.26}). For the sake of brevity, one can check  
\begin{eqnarray*}
s_b Q_{ab} &=& - i \int d^2x\,\Big[\bar B \cdot \partial^0\Big(B + \bar B + i g C \times \bar C\Big)\Big]  \nonumber\\
&+& \int d^2x\,\Big[g \Big(\big(B + \bar B + i g C \times \bar C\big) \times \beta \Big) \cdot D^0\beta
- g D^0\Big(\big(B + \bar B + i g C \times \bar C\big) \times \beta \Big) \cdot \bar \beta\Big] \nonumber\\
&-& i\int d^2x \Big[\big(R + \bar R + i g C \times \bar \beta 
+ i g \bar C \times \beta\big) \cdot D^0\big(R + i g C \times \bar \beta\big)\Big],\nonumber\\
\end{eqnarray*}
\begin{eqnarray}
s_{ab} Q_b &=&  i \int d^2x\,\Big[B \cdot \partial^0\Big(B + \bar B + i g C \times \bar C\Big)\Big]  \nonumber\\
&-& \int d^2x\,\Big[g \Big(\big(B + \bar B + i g C \times \bar C\big) \times \bar \beta \Big) \cdot D^0 \beta
- g D^0\Big(\big(B + \bar B + i g C \times \bar C\big) \times \bar \beta \Big) \cdot  \beta\Big] \nonumber\\
&+& i\int d^2x \Big[\big(R + \bar R + i g C \times \bar \beta 
+ i g \bar C \times \beta\big) \cdot D^0\big(\bar R + i g \bar C \times  \beta\big)\Big].
\end{eqnarray}
It is clear from the above expressions that $s_b Q_{ab} = 0$ and $s_{ab} Q_b = 0$ if and only if  CF conditions (\ref{3.26}) are satisfied. 
As a consequence, the (anti-)BRST charges are anticommuting only on the constrained hypersurface defined by the 
CF conditions (\ref{3.26}).

\section{Ghost scale symmetry and BRST algebra}

The Lagrangian densities (\ref{4.2}), in addition to the (anti-)BRST symmetry transformations, also respect the continuous ghost 
scale symmetry $(s_g)$. These symmetry transformations  are given as follows      
\begin{eqnarray}
&&C \to e^{+\Omega}\,C, \qquad \bar C \to e^{-\Omega}\,\bar C, \qquad \beta \to e^{+\Omega}\,\beta, 
\qquad \bar \beta \to e^{-\Omega}\,\bar \beta, \nonumber\\
&& \big(A_\mu, \phi_\mu, \rho, B, \bar B, R, \bar R\big) \to e^0 \big(A_\mu, \phi_\mu, \rho, B, \bar B, R, \bar R\big)
\end{eqnarray}
where $\Omega$ is the global scale parameter. The numbers ($\pm1, 0$) in the exponential of the above
transformations stand for ghost numbers of the corresponding fields. For instance, the ghost fields $(C, \beta)$ carry ghost number $(+1)$
and anti-ghost fields $(\bar C, \bar \beta)$ have ghost number ($-1$). The rest (bosonic) fields have ghost number zero. 
The infinitesimal version of the above continuous transformation is given by  
\begin{eqnarray}
&& s_g C = + \Omega \,C, \qquad s_g \bar C = -  \Omega \,\bar C, \qquad s_g \beta = +  \Omega \,\beta, 
\qquad s_g \bar \beta = -  \Omega \bar \beta, \nonumber\\
&&  s_g\big(A_\mu, \phi_\mu, \rho, B, \bar B, R, \bar R\big) =0. \label{6.2}
\end{eqnarray}
It is straightforward to check that under the above continuous ghost scale symmetry transformations (\ref{6.2}) both the Lagrangian densities
remain invariant (i.e. $s_g{\cal L}_b = s_g {\cal L}_{\bar b} = 0$). As a consequence, the existence of ghost scale symmetry 
leads to the following Noether's conserved current ($J^\mu_g$) and charge ($Q_g$):   
\begin{eqnarray}
J^\mu_g &=& i \Big[\bar C \cdot (D^\mu C) - (\partial^\mu \bar C) \cdot C 
+ \bar \beta \cdot (D^\mu \beta) -  (D^\mu \bar \beta) \cdot \beta \Big], \nonumber\\
Q_g &=& i \int d^2x \,\Big[\bar C \cdot (D^0 C) - (\partial^0 \bar C) \cdot C 
+ \bar \beta \cdot (D^0 \beta) -  (D^0 \bar \beta) \cdot \beta \Big].
\end{eqnarray}
The conservation law $(\partial_\mu J^\mu_g = 0)$ can be proven by exploiting the E-L equations of motion (\ref{5.2}).
The ghost charge $Q_g$ also turns out to be the generator of the ghost scale symmetry transformations (\ref{6.2}). For instance, one can check that
$s_g C = - i [C, \, \Omega\,Q_g] = +\Omega\, C.$

The above ghost charge $Q_g$ together with the nilpotent (anti-)BRST charges $Q_{(a)b}$ obey a standard BRST algebra.
In operator form, this algebra can be given as follow  
\begin{eqnarray}
&& Q^2_b = 0, \qquad Q^2_{ab} = 0, \qquad \big\{Q_b,\; Q_{ab}\big\} = Q_b\,Q_{ab} + Q_{ab}\, Q_b = 0,\nonumber\\
&& i \big[Q_g, \,Q_b\big] = + Q_b, \qquad i \big[Q_g, Q_{ab}\big] = - Q_{ab}, \qquad Q^2_g \ne 0. \label{6.4}
\end{eqnarray}

Let us consider a state $|\psi \rangle _n$, in the quantum Hilbert space of states, such that the ghost number of the state 
is defined in the following manner
\begin{eqnarray}
 i Q_g |\psi \rangle _n =  n |\psi \rangle _n ,
\end{eqnarray}
where $n$ is the ghost number of the state $|\psi \rangle _n$. Now, it is easy to check, with the help of above algebra (\ref{6.4}),
that following relationships holds 
\begin{eqnarray}
&& i Q_g Q_b |\psi \rangle _n = (n + 1) Q_b |\psi \rangle _n , \nonumber\\
&& i Q_g Q_{ab} |\psi \rangle _n = (n - 1) Q_{ab} |\psi \rangle _n , 
\end{eqnarray}
which shows that the BRST charge $Q_b$ increases the ghost number by one unit when it operates on a quantum state
whereas the anti-BRST charge $Q_{ab}$ decreases it by one unit. In other words, we can say that the (anti-)BRST 
charge carry the ghost numbers $(\mp 1)$, respectively. A careful look at the expressions of the (anti-)BRST and 
ghost charges, where the ghost numbers of the fields are concerned, also reveal the same observations.

\section{Role of auxiliary field: A bird's-eye view}
In this section we provide a brief synopsis about few striking similarities and some glaring differences among the 3D non-Abelian JP model, 
4D  topologically massive non-Abelian 2-form gauge theory \cite{Malik:2010gu,Kumar:2010kd} and the 4D modified gauge invariant Proca theory  
in the realm of well-known St{\"u}ckelberg formalism (see, e.g. \cite{Ruegg:2003ps} for details).

\subsection{Jackiw-Pi model}
It is interesting to note that, if we make the following substitution
\begin{eqnarray}
\phi_\mu \;\longrightarrow \; \phi_\mu + D_\mu \rho, \label{7.1}
\end{eqnarray}
in our starting Lagrangian density (\ref{2.1}), the 2-form $G_{\mu\nu}$ and 
mass term re-defined as
\begin{eqnarray}
G_{\mu\nu}& \longrightarrow & G_{\mu\nu} - g \big(F_{\mu\nu} \times \rho \big),\nonumber\\
\frac{m}{2}\,\varepsilon^{\mu\nu\eta}\,F_{\mu\nu} \cdot \phi_\eta & \longrightarrow &
 \frac{m}{2}\,\varepsilon^{\mu\nu\eta}\,F_{\mu\nu} \cdot \phi_\eta
+ \partial_\eta \Big[\frac{m}{2}\, \varepsilon^{\mu\nu\eta}\,F_{\mu\nu} \cdot \rho \Big] 
- \frac{m}{2}\,\varepsilon^{\mu\nu\eta}\,\big(D_\eta F_{\mu\nu}\big) \cdot \rho. \qquad
\end{eqnarray}
In  the above,  the term $\frac{m}{2}\,\varepsilon^{\mu\nu\eta}\,\big(D_\eta F_{\mu\nu}\big) \cdot \rho$ is zero 
 due to the validity of the well-known Bianchi identity 
$(D_\mu F_{\nu\eta} + D_\nu F_{\eta\mu} + D_\eta F_{\mu\nu} = 0)$.  
Therefore, the mass term remains invariant, modulo a total spacetime derivative, under the re-definition (\ref{7.1}). 
As a consequence, the modified Lagrangian density, modulo a total spacetime derivative, is given by
\begin{eqnarray}
\tilde{\cal L}_0 = - \frac{1}{4}\, F_{\mu\nu}\cdot F^{\mu\nu} 
- \frac{1}{4}\, G_{\mu\nu}\cdot G^{\mu\nu} + \frac{m}{2}\, \varepsilon^{\mu\nu\eta}\,F_{\mu\nu}\cdot \phi_\eta. \label{7.3}
\end{eqnarray} 
 It is clear that the auxiliary field $\rho$ is completely eliminated from the above Lagrangian density. 
We point out that, even though, Lagrangian density (\ref{7.3}) respects the YM gauge transformations (\ref{2.2}) 
but it fails to respect the NYM gauge transformations (\ref{2.3}). 
The similar observation can also be seen in the case of 4D topologically massive non-Abelain 2-form gauge theory as well as in the 
4D modified gauge invariant version of Proca theory.

\subsection{4D massive non-Abelian 2-form gauge theory}
The Lagrangian density for the 4D massive non-Abelian 2-form gauge theory is given by (see, for details \cite{Kumar:2011zi,Malik:2010gu,Kumar:2010kd})
\begin{eqnarray}
{\cal L} = - \frac{1}{4}\, F_{\mu\nu}\cdot F^{\mu\nu} 
+ \frac{1}{12}\, H_{\mu\nu\eta}\cdot H^{\mu\nu\eta} + \frac{m}{4}\, \varepsilon^{\mu\nu\eta\kappa}\,B_{\mu\nu}\cdot F_{\eta\kappa}, \label{7.4}
\end{eqnarray} 
where 3-form $H_{\mu\nu\eta} = D_\mu B_{\nu\eta} + D_\nu B_{\eta\mu} + D_\eta B_{\mu\nu} + g(F_{\mu\nu} \times K_\eta)
+ g(F_{\nu\eta} \times K_\mu)+ g (F_{\eta\mu} \times K_\nu)$ is the field strength tensor corresponding to the 2-form gauge field
$B_{\mu\nu}$ and the 2-form field strength tensor $F_{\mu\nu} = \partial_\mu A_\nu - \partial_\nu A_\mu - g (A_\mu \times A_\nu)$ 
corresponds to the 1-form gauge field $A_\mu$.  The coupling constant is represented by $g$ and $D_\mu$ is the covariant derivative. 
The auxiliary field $K_\mu$ is the compensating field.
This Lagrangian density respects the two types of gauge transformations -- the scalar gauge transformation $(\tilde \delta_1)$ 
and vector gauge transformation $(\tilde \delta_2)$, namely; \cite{Kumar:2011zi,Malik:2010gu,Kumar:2010kd}
\begin{eqnarray}
&&\tilde \delta_1 A_\mu = D_\mu \Omega, \qquad \tilde \delta_1 B_{\mu\nu} = - g(B_{\mu\nu} \times \Omega),
 \qquad \tilde \delta_1 K_\mu = - g(K_\mu \times \Omega), \nonumber\\
&& \tilde \delta_2 A_\mu =0, \qquad \tilde \delta_2 B_{\mu\nu} = - (D_\mu \Lambda_\nu - D_\nu \Lambda_\mu), 
\qquad \tilde \delta_2 K_\mu = - \Lambda_\mu, \label{7.5}
\end{eqnarray}   
where $\Omega (x)$ and $\Lambda_\mu (x)$ are the local scalar and vector gauge parameters, respectively.
We note that if we re-define the $B_{\mu\nu}$ field as
\begin{eqnarray}
B_{\mu\nu} \;\longrightarrow \; B_{\mu\nu} + (D_\mu K_\nu - D_\nu K_\mu),
\end{eqnarray}
the  3-form field strength tensor $H_{\mu\nu\eta}$ and the mass term modify as follows 
\begin{eqnarray}
H_{\mu\nu\eta} & \longrightarrow&  \tilde H_{\mu\nu\eta}\;=\; D_\mu B_{\nu\eta} + D_\nu B_{\eta\mu} + D_\eta B_{\mu\nu}, \nonumber\\
\frac{m}{4}\,\varepsilon^{\mu\nu\eta\kappa}\,B_{\mu\nu}  \cdot F_{\eta\kappa} &\longrightarrow& 
 \frac{m}{4}\,\varepsilon^{\mu\nu\eta\kappa}\,B_{\mu\nu}  \cdot F_{\eta\kappa}
+ \partial_\mu \Big[\frac{m}{2}\, \varepsilon^{\mu\nu\eta\kappa}\,  K_\nu \cdot F_{\eta\kappa} \Big] \nonumber\\
&& \qquad \qquad \qquad \qquad  -\; \frac{m}{2}\,\varepsilon^{\mu\nu\eta\kappa}\,K_\nu \cdot \big(D_\mu F_{\eta\kappa}\big). 
\end{eqnarray}
and the compensating auxiliary vector field $K_\mu$ disappears from the Lagrangian density (\ref{7.4}). Furthermore, the mass term 
$\displaystyle \frac{m}{4}\,\varepsilon^{\mu\nu\eta\kappa}\,B_{\mu\nu}  \cdot F_{\eta\kappa}$ remains intact modulo a total spacetime derivative.
Thus, the modified Lagrangian density can be given in the following manner (modulo a total spacetime derivative)
\begin{eqnarray}
\tilde {\cal L} = - \frac{1}{4}\, F_{\mu\nu}\cdot F^{\mu\nu} 
+ \frac{1}{12}\, {\tilde H}_{\mu\nu\eta}\cdot \tilde{H}^{\mu\nu\eta} 
+ \frac{m}{4}\, \varepsilon^{\mu\nu\eta\kappa}\,B_{\mu\nu}\cdot F_{\eta\kappa}. \label{58}
\end{eqnarray}
Clearly, the above Lagrangian density is no longer invariant under the vector gauge transformation  
even though it respects the scalar gauge transformations [cf. (\ref{7.5})].

It is clear form the above discussions that both the above models (i.e. JP model and 4D massive non-Abelian 2-form gauge theory) 
are very similar to each other in the sense that under the re-definitions of the
fields $\phi_\mu$ and $B_{\mu\nu}$ the auxiliary fields $\rho$ and $K_\mu$ are eliminated from their respective models. 
As a result, the modified Lagrangian densities (\ref{7.3}) and (\ref{58}) do not respect the     
symmetry transformations $(\delta_2)$ and $(\tilde \delta_2)$, respectively. Thus, the auxiliary fields $\rho$ and $K_\mu$ are required 
in their respective models so that these models respect both the gauge symmetry transformations [cf. (\ref{2.2}), (\ref{2.3}) and (\ref{7.5})].

\subsection{Modified version of Abelian Proca theory} 
The above key observations can also be seen in the case of modified gauge invariant Abelian Proca theory.
The gauge invariant Lagrangian density of this model is as follows \cite{Ruegg:2003ps}
\begin{eqnarray}
{\cal L}_s = - \frac{1}{4}\, F_{\mu\nu}\,F^{\mu\nu} + \frac{m^2}{2}\, A_\mu \,A^\mu 
+ \frac{1}{2}\,\partial_\mu \phi\,\partial^\mu \phi + m A_\mu\, \partial^\mu \phi, \label{7.8}
\end{eqnarray} 
 where $F_{\mu\nu} = \partial_\mu A_\nu - \partial_\nu A_\mu$ is the field strength tensor corresponding to $A_\mu$, 
$\phi$ is the St{\"u}ckelberg field and $m$ represents the mass of the photon field $A_\mu$. Under the following 
 local gauge transformations  
\begin{eqnarray}
\delta_{(gt)}A_\mu = \partial_\mu \chi(x), \qquad \delta_{(gt)} \phi = - m\,\chi(x), \label{7.9}
\end{eqnarray}
the Lagrangian density (\ref{7.8}) remains invariant. Here $\chi(x)$ is the local gauge transformation parameter.
It can be checked that under the following re-definition
\begin{eqnarray}
A_\mu \;\longrightarrow\; A_\mu - \frac{1}{m}\,\partial_\mu \phi,
\end{eqnarray} 
the St{\"u}ckelberg field $\phi$ completely disappears from the Lagrangian density (\ref{7.8}). As a consequence, the resulting Lagrangian density does not respect 
the above gauge transformations (\ref{7.9}).

The above observation is very similar to the JP model and the massive  non-Abelian 2-form gauge theory. As a consequence, 
the field $\rho$ (in JP model) and $K_\mu$ (in massive non-Abelian 2-form theory) are like the St{\"u}ckelberg field. However, 
the key difference is that these St{\"u}ckelberg like fields (i.e. $\rho$ and $K_\mu$) are auxiliary fields in their respective models
whereas, in the modified gauge invariant Proca theory, the St{\"u}ckelberg field $\phi$ is dynamical in nature.

\section {Conclusions}
In our present investigation, we have derived the off-shell nilpotent and absolutely anticommuting (anti-)BRST symmetry transformations 
corresponding to the combined YM and NYM symmetries of the JP model. For this purpose, we have utilized the power and strength of augmented
superfield approach. The derivation of proper (anti-)BRST symmetries for the auxiliary field $\rho$ is one of the main findings of our 
present endeavor. These (anti-)BRST symmetry transformations corresponding to the auxiliary field $\rho$ can neither be generated 
from the conserved (anti-)BRST charges nor deduced by the requirement of nilpotency and/or absolute anticommutativity of the (anti-)BRST 
symmetry transformations.

One of the main features of the superfield formalism is the derivation of CF conditions which, in turn, ensure the absolutely anticommutativity
of (anti-) BRST symmetry transformations. The CF conditions, a hallmark of any non-Abelian 1-form gauge theories \cite{Curci:1976ar},
appear naturally within the framework of superfield formalism and also have connections with gerbes \cite{Bonora:2007hw}. In our present case of 
combined YM and NYM symmetries of JP model, there exist {\it two} CF conditions 
(cf. (\ref{3.26})). This is in contrast to the YM symmetries case where there exist only {\it one} CF condition \cite{Gupta:2011cta}
and in NYM symmetries case, {\it no} CF condition was observed \cite{Gupta:2012ur}. Moreover, these CF conditions have played a central role 
in the derivation of coupled Lagrangian densities (cf. Section 4).

Furthermore, we have obtained a set of coupled Lagrangian densities which respect the above mentioned (anti-)BRST symmetry transformations. 
The ghost sector of these coupled Lagrangian densities is also endowed with another continuous symmetry - the ghost symmetry.  
We have exploited this symmetry to derive the conserved ghost charge. Moreover, we have pointed out the standard BRST
algebra obeyed by all the conserved charges of the underlying theory.

At the end, we have provided a bird's-eye view on the role of auxiliary field in the context of various massive models. For this purpose,
we have taken three different cases of 3D JP model, 4D massive non-Abelian 2-form gauge theory and the 4D modified version of 
Abelian Proca theory. We have shown that the field $\rho$ (in JP model) and $K_\mu$ (in massive non-Abelian 2-form theory) 
are like St{\"u}ckelberg field ($\phi$) of Abelian Proca model. However, $\rho$ and $K_\mu$ are auxiliary fields whereas $\phi$ is 
dynamical, in their respective models. Finally, we capture the (anti-)BRST invariance of the coupled Lagrangian densities
(cf. (\ref{4.2})), nilpotency and absolute anticommutativity of (anti-)BRST charges (cf. (\ref{5.5})) within the framework 
of superfield approach.

\section*{Acknowledgments}
The research work of SG is supported by Conselho Nacional de Desenvolvimento Cient\'{i}fico e Tecnol\'{o}gico (CNPq) grant 151112/2014-2.

\appendix 
\section{(Anti-)BRST invariance, nilpotency and anticommutativity: Superfield approach}
It is interesting to point out that the super expansions (\ref{3.5}), (\ref{3.15}) and (\ref{3.22}) can be expressed in terms 
of the translations of the corresponding superfields along the Grassmannian directions of the $(3,2)$-dimensional 
supermanifold, as 
\begin{eqnarray}
&& s_b \Psi (x)  =  \frac {\partial}{\partial \bar \theta} \, \tilde \Psi^{(h)}(x,\theta,\bar\theta)\Big|_{\theta = 0}, \qquad \quad
s_{ab} \Psi (x)  =   \frac {\partial}{\partial \theta} \, \tilde \Psi^{(h)}(x,\theta,\bar\theta)\Big|_{\bar \theta = 0},\nonumber\\
&&s_b\, s_{ab} \Psi (x)  =  \frac {\partial}{\partial \bar \theta} \, \frac {\partial}{\partial \theta} 
\, \tilde \Psi^{(h)} (x,\theta,\bar\theta), \label{b1}
\end{eqnarray}
where $\Psi (x)$ is any generic field of the underlying 3D theory and $\Psi^{(h)} (x,\theta,\bar\theta)$ is the corresponding superfield 
obtained after the application of HC. The above expression captures the off-shell nilpotency of the (anti-)BRST symmetries because of the 
properties of Grassmannian derivatives, i.e. $\partial_\theta^2 = \partial_{\bar \theta}^2 = 0$.  
Moreover, the anticommutativity property of the (anti-)BRST symmetry transformations is also clear from the expansions (\ref{3.5}), (\ref{3.15}) 
and (\ref{3.22}), in the following manner
\begin{eqnarray}
\Big(\frac {\partial}{\partial \theta} \; \frac {\partial}{\partial \bar \theta} +
\frac {\partial}{\partial \bar \theta} \; \frac {\partial}{\partial \theta} \Big) \; \tilde \Psi^{(h)} 
(x, \theta, \bar \theta )  =  0. \label{b2}
\end{eqnarray}
Thus, the expressions (\ref{b1}) and (\ref{b2}) provide the geometrical interpretations for the (anti-) BRST symmetry 
transformations in terms of the translational generators $(\partial_\theta, \partial_{\bar\theta})$ along the Grassmannian 
directions of the $(3, 2)$-dimensional supermanifold. 

Furthermore, the nilpotency of the (anti-)BRST charges can also be realized, within the framework of superfield formalism, in the 
following manner 
\begin{eqnarray}
Q_b &=& \frac {\partial}{\partial \bar \theta} \, \int d^2 x \,
\Big[B(x) \cdot \tilde {\cal A}_0^{(h)} (x, \theta, \bar\theta) + i \, {\dot{\tilde {\bar {\cal F}}}}^{(h)}(x, \theta, \bar\theta)
\cdot \tilde {\cal F}^{(h)} (x, \theta, \bar\theta)\nonumber\\
&+& \Big(R(x)+ ig \tilde {\cal F}^{(h)}(x, \theta, \bar\theta) \times {\tilde {\bar \beta}}^{(h)}(x, \theta, \bar\theta) \Big) 
\cdot \tilde\Phi^{(h)}_0 (x, \theta, \bar\theta) \nonumber\\
&+& i {\tilde {\cal D}}_0{\tilde {\bar \beta}}^{(h)}
(x, \theta, \bar\theta) \cdot \tilde \beta^{(h)} (x, \theta, \bar\theta)\Big] \bigg|_{\theta = 0} \nonumber\\
&\equiv& \int d^2 x  \int d \bar\theta \; 
\Big[B(x) \cdot \tilde {\cal A}_0^{(h)} (x, \theta, \bar\theta) + i \; {\dot{\tilde {\bar {\cal F}}}}^{(h)}(x, \theta, \bar\theta)
\cdot \tilde {\cal F}^{(h)} (x, \theta, \bar\theta ) \nonumber\\
&+& \Big(R(x)+ ig \tilde {\cal F}^{(h)}(x, \theta, \bar\theta) \times {\tilde {\bar \beta}}^{(h)}(x, \theta, \bar\theta) \Big) 
\cdot \tilde\Phi^{(h)}_0 (x, \theta, \bar\theta) \nonumber\\
 &+& i {\tilde {\cal D}}_0{\tilde {\bar \beta}}^{(h)}
(x, \theta, \bar\theta) \cdot \tilde \beta^{(h)} (x, \theta, \bar\theta)\Big] \bigg|_{\theta = 0}. 
\end{eqnarray}
This, in turn, implies  
\begin{eqnarray}
\frac {\partial}{\partial \bar\theta} \; Q_b \bigg|_{\theta = 0}  =  0 
\quad \Longrightarrow \quad Q_b^2  =  0,
\end{eqnarray} 
because of the nilpotency property of the Grassmannian derivative (i.e. $\partial_{\bar \theta}^2 = 0$). 
It is interesting to point out that the nilpotency of above BRST charge $(Q_b)$, when written in ordinary 3D spacetime,   
\begin{eqnarray}
Q_b &=& \int d^2 x\, s_b \ \Big[B(x) \cdot A_0 (x) + i \; \dot{\bar C} (x) \cdot C(x) 
+ \Big(R(x) + i g C(x) \times \bar \beta(x)\Big)\cdot \phi_0(x) \nonumber\\
&+& i D_0 \bar \beta(x) \cdot \beta(x) \Big],
\end{eqnarray}
is straightforward and encoded in the nilpotency property $(s_b^2 = 0)$ of the BRST transformations 
$(s_b)$. In other words, $s_b Q_b = -i \{Q_b, Q_b\} = 0$ is true due to above mentioned reason. 
Moreover, using the CF-conditions, there is yet another way to express the above BRST charge where nilpotency is quite clear, 
as can be seen from the following expression
\begin{eqnarray}
Q_b &=& i \frac {\partial} {\partial \bar \theta} \, \frac {\partial} {\partial  \theta} \int
d^2 x  \Big[ \tilde {\cal A}_0^{(h)} (x, \theta, \bar \theta) \cdot \tilde {\cal F}^{(h)} 
(x, \theta, \bar \theta)  +  \tilde \Phi_0^{(h)} (x, \theta, \bar \theta) \cdot \tilde \beta^{(h)} 
(x, \theta, \bar \theta)\Big] \nonumber\\
 &\equiv& i \int d^2 x \, s_b\, s_{ab} \, \Big[\, A_0 (x) \cdot C(x) + \phi_0(x) \cdot \beta(x)\Big].
\end{eqnarray}
This is true only on the constrained surface spanned by CF conditions. Similarly, we can express the anti-BRST charge $(Q_{ab})$
in the following two different ways: 
\begin{eqnarray}
Q_{ab}  &= & - \frac {\partial}{\partial \theta} \, \int d^2 x \,
\Big[\bar B(x) \cdot \tilde {\cal A}_0^{(h)} (x, \theta, \bar\theta) + i \, \dot{\tilde{{\cal F}}}^{(h)}(x, \theta, \bar\theta)
\cdot {\tilde {\bar {\cal F}}}^{(h)} (x, \theta, \bar\theta)\nonumber\\
&-& \Big(\bar R(x)+ ig {\tilde {\bar {\cal F}}}^{(h)}(x, \theta, \bar\theta) \times {\tilde {\beta}}^{(h)}(x, \theta, \bar\theta) \Big) 
\cdot \tilde\Phi^{(h)}_0(x, \theta, \bar\theta)\nonumber\\
&-& i {\tilde {\cal D}}_0{\tilde {\beta}}^{(h)}
(x, \theta, \bar\theta) \cdot {\tilde {\bar \beta}}^{(h)} (x, \theta, \bar\theta)\Big] \bigg|_{\theta = 0} \nonumber\\
& \equiv & i \frac {\partial}{\partial \theta} \frac {\partial}{\partial \bar \theta} 
\int d^2 x   \Big[ \tilde {\cal A}_0^{(h)} (x, \theta, \bar\theta) \cdot 
{\tilde {\bar {\cal F}}}^{(h)} (x, \theta, \bar\theta) 
+ \tilde \Phi_0^{(h)} (x, \theta, \bar \theta) \cdot {\tilde {\bar \beta}}^{(h)} (x, \theta, \bar \theta)\Big]. \label{A7}
\end{eqnarray}
In the above, the second expression is valid on the constrained hypersurface parametrized by the CF conditions. The 
nilpotency of anti-BRST charge (i.e. $Q_{ab}^2 = 0$) is assured by the nilpotency $(\partial_\theta^2 = 0)$ of the Grassmannian 
derivative $\partial_\theta$, as described below 
\begin{eqnarray}
\frac {\partial}{\partial \theta} \; Q_{ab} \bigg|_{\bar \theta = 0}  =  0 
\quad \Longrightarrow \quad Q_{ab}^2  =  0.
\end{eqnarray} 
In 3D ordinary space, the above expression (\ref{A7}) can be written in the following fashion
\begin{eqnarray}
Q_{ab} &=& - \int d^2 x \; s_{ab}\;  \Big [ \bar B(x) \cdot A_0(x) + i \; \dot C (x) \cdot \bar C (x) 
+ \Big(\bar R(x) + i g \bar C(x) \times \beta(x)\Big)\cdot \phi_0(x) \nonumber\\
&+& i D_0 \beta(x) \cdot \bar \beta(x)\Big] \nonumber\\
&\equiv& - i\, \int s_{ab} s_b \,\Big[\, A_0 (x) \cdot \bar C(x) + \phi_0(x) \cdot \bar \beta(x)\Big].
\end{eqnarray}
Here, the nilpotency of the anti-BRST charge lies in the equation $s_{ab} Q_{ab} = - i \{Q_{ab}, Q_{ab} \} = 0$ because 
of the fact that $s_{ab}^2 = 0$.

Furthermore, in order to prove the (anti-)BRST invariance of the coupled Lagrangian densities, within the framework of 
superfield formalism, we first generalize our starting Lagrangian density (${\cal L}_0$)
onto the $(3,2)$-dimensional supermanifold, as follows
\begin{eqnarray}
{\cal L}_0 \to \tilde {\cal L}_0 &=& - \frac{1}{4}\, \tilde {\cal F}^{\mu\nu (h)} 
\cdot \tilde {\cal F}_{\mu\nu}^{(h)}  - \;\frac{1}{4}\, \Big[\tilde {\cal G}^{\mu\nu (h)} 
+ g \, \tilde {\cal F}^{\mu\nu (h)} \times \tilde \rho^{(h)} \Big] 
\cdot \Big[\tilde {\cal G}_{\mu\nu}^{(h)} + g \, \tilde {\cal F}_{\mu\nu}^{(h)} \times \tilde \rho^{(h)} \Big] \nonumber\\ 
&+& \frac {m}{2}\,\varepsilon^{\mu\nu\eta} \, \tilde {\cal F}_{\mu\nu}^{(h)} \cdot \tilde \Phi_\eta^{(h)}.
\end{eqnarray}
This Lagrangian density ($\tilde {\cal L}_0$) is free from the Grassmannian variables (cf. Section 3, for details). Therefore, the followings are  true
\begin{eqnarray}
\frac {\partial}{\partial \bar \theta} \; \tilde {\cal L}_0\bigg|_{\theta = 0}  =  0, \qquad
\frac {\partial}{\partial  \theta} \; \tilde {\cal L}_0\bigg|_{\bar \theta = 0} =  0,  
\end{eqnarray}
which captures the (anti-)BRST invariance of the starting Lagrangian density ${\cal L}_0$. Similarly, we
can also generalize the coupled Lagrangian densities (\ref{4.2}) onto the $(3,2)$-dimensional supermanifold 
in the following manner
\begin{eqnarray}
{\cal L}_{\bar b} \longrightarrow \tilde {\cal L}_{\bar b} & =& \tilde {\cal L}_0 - \frac {\partial} 
{\partial \theta}\, \frac {\partial} {\partial \bar \theta}\,  \Big[ \frac {i} {2} \, \tilde {\cal A}_\mu^{(h)} 
\cdot \tilde {\cal A}^{\mu (h)} + \tilde {\cal F}^{(h)} \cdot {\tilde {\bar {\cal F}}}^{(h)} 
+ \frac {i} {2} \, \tilde \Phi_\mu^{(h)} \cdot \tilde \Phi^{\mu (h)}
 + \frac{1}{2}\, \tilde  \beta^{(h)} \cdot {\tilde {\bar \beta}}^{(h)} \Big],\nonumber\\
&&\nonumber\\
{\cal L}_{b} \longrightarrow \tilde {\cal L}_{b} & = & \tilde {\cal L}_0 + \frac {\partial} 
{\partial \bar \theta}\, \frac {\partial} {\partial  \theta}\,  \Big [ \frac {i} {2} \, \tilde {\cal A}_\mu^{(h)} 
\cdot \tilde {\cal A}^{\mu (h)} + \tilde {\cal F}^{(h)} \cdot {\tilde {\bar {\cal F}}}^{(h)} 
+ \frac {i}{2} \, \tilde \Phi_\mu^{(h)} \cdot \tilde \Phi^{\mu (h)} 
+ \frac{1}{2}\, \tilde  \beta^{(h)} \cdot {\tilde {\bar \beta}}^{(h)}  \Big].\nonumber\\
\end{eqnarray}
Now, the (anti-)BRST invariance of the above coupled Lagrangian densities is straightforward 
because of the fact $(\partial_\theta^2 = \partial_{\bar \theta}^2 = 0)$. Thus, we have
\begin{eqnarray}
\frac {\partial}{\partial \theta}\;  \tilde{\cal L}_{\bar b}\bigg|_{\bar \theta = 0}  = 0,\qquad
\frac {\partial}{\partial \bar \theta}\;  \tilde{\cal L}_b \bigg|_{\theta = 0} = 0,
\end{eqnarray}
which imply the (anti-)BRST invariance of the coupled Lagrangian densities within the framework of 
superfield formalism.

\end{document}